\documentclass[a4paper,6pt,intlimits]{iopart}
\expandafter\let\csname equation*\endcsname\relax
\expandafter\let\csname endequation*\endcsname\relax
\usepackage[tbtags]{amsmath}
\usepackage{amsthm,amssymb,dsfont,mathrsfs,bbm,xfrac,tabularx,booktabs}
\usepackage{cite}
\usepackage{mathtools}
\usepackage{multirow}
\usepackage{upgreek}
\usepackage[usenames,dvipsnames]{xcolor}
\usepackage[caption=false]{subfig}
\usepackage[english]{babel}
\usepackage{pgfplotstable}
\usepackage{hyperref}
\hypersetup{pdfauthor={Lucibello Malatesta Parisi Sicuro},pdftitle={Fractional matching},%
            colorlinks, linktocpage=true, pdfstartpage=3, pdfstartview=FitV,%
    breaklinks=true, pdfpagemode=UseNone, pageanchor=true, pdfpagemode=UseOutlines,%
    plainpages=false, bookmarksnumbered, bookmarksopen=true, bookmarksopenlevel=1,%
    hypertexnames=true, pdfhighlight=/O,%
    urlcolor=orange, linkcolor=blue, citecolor=red, 
        }

\usepackage{lmodern}
\usepackage[T1]{fontenc}
\usepackage[cal=boondoxo]{mathalfa}

\catcode`,\active

\catcode`\,12

\newcommand{\bQ}{\boldsymbol{\mathsf Q}}
\newcommand{\mE}{\mathcal{E}}
\newcommand{\mV}{\mathcal{V}}
\newcommand{\He}{\mathrm{He}}

\newcommand{\R}{\mathds{R}}

\newcommand{\N}{\mathds{N}}
\DeclareMathOperator{\dd}{d}

\allowdisplaybreaks
\begin{document}
\title[The Random Fractional Matching Problem]{The Random Fractional Matching Problem}
\date{\today}

\author{Carlo Lucibello}\ead{carlo.lucibello@polito.it}
\address{Dept. of Applied Science and Technology, Politecnico di Torino, and Italian Institute for Genomic Medicine, Torino, Italy}
\author{Enrico M. Malatesta}\ead{enrico.m.malatesta@gmail.com}
\address{Dipartimento di Fisica, University of Milan and INFN, via Celoria 16, I-20133 Milan, Italy}
\author{Giorgio Parisi}\ead{giorgio.parisi@roma1.infn.it}
\address{Dipartimento di Fisica, Sapienza Universit\`a di Roma, INFN Sezione di Roma1, and CNR-NANOTEC UOS Roma, P.le A. Moro 2, I-00185, Rome, Italy}
\author{Gabriele Sicuro}\ead{gabriele.sicuro@roma1.infn.it}
\address{Dipartimento di Fisica, Sapienza Universit\`a di Roma, P.le A. Moro 2, I-00185, Rome, Italy}

\begin{abstract}
 We consider two formulations of the random-link fractional matching problem, a relaxed version of the more standard random-link (integer) matching problem. In one formulation, we allow each node to be linked to itself in the optimal matching configuration. In the other one, on the contrary, such a link is forbidden. Both problems have the same asymptotic average optimal cost of the random-link matching problem on the complete graph. Using a replica approach and previous results of W{\"a}stlund \cite{Wastlund2010}, we analytically derive the finite-size corrections to the asymptotic optimal cost. We compare our results with numerical simulations and we discuss the main differences between random-link fractional matching problems and the random-link matching problem. 
\end{abstract}
\section{Introduction}
Random combinatorial optimization problems (\textsc{Rcop}s) are optimization problems in which some parameters of the cost function are random variables. \textsc{Rcop}s live therefore at the intersection of mathematical optimization theory and probability theory, and typically the average properties of the solution of the problem are of primary interest. Surprisingly enough, methods developed for the study of disordered systems, such as the replica approach and the cavity method, have proven to be powerful tools for the study of \textsc{Rcop}s, they have been able to provide precise analytical predictions \cite{Kirkpatrick1983,Mezard1985,mezard1987spin,mezard2009information,Bapst2013}, and they inspired new algorithms for their solution \cite{mezard2009information,Braunstein2005,Braunstein2006,Altarelli2011,Bapst2013}. Indeed, the search for an optimal solution in a \textsc{Rcop} can be thought as the search for the ground state of a constrained disordered system, whose energy coincides with the cost function appearing in the original \textsc{Rcop}. Despite the fact that many of the results obtained by this analogy were originally non-rigorously derived, they were later confirmed by rigorous mathematical proofs \cite{Aldous1992,Aldous2001,Nair2003,Linusson2004,Wastlund2008,Wastlund2010}.

In this paper we will consider an example of \textsc{Rcop} that belongs to a well studied class of combinatorial optimization problems, namely the class of \textit{matching problems}. Minimum matching problems are combinatorial optimization problems that can be defined on a graph $\mathcal G=(\mV,\mE)$ having vertex set $\mathcal V$ and edge set $\mathcal E\subseteq\mathcal V\times\mathcal V$, both of finite cardinality. The graph is said to be {weighted} if a weight $w_e\in\R$ is associated to each edge $e\in\mE$. A (minimum) matching is therefore a subset $\mathcal M\subseteq\mE$ of edges that do not share a common vertex, such that the cardinality $|\mathcal M|$ is maximal and the cost of the matching, defined as $\sum_{e\in\mathcal M}m_ew_e$, is minimum. If $\mathcal M$ matches all vertices in $\mV$, the matching is said to be {perfect} \cite{lovasz2009matching,diestel2012graph}. Once the weighted graph is assigned, the solution of a specific instance of the matching problem on a graph can be found quite easily using efficient algorithms available in the literature \cite{papadimitriou1998combinatorial,Kuhn1955a,Munkres1957,Jonker1987,mezard2009information}. 

\begin{figure}
\centering
 \includegraphics[width=\textwidth]{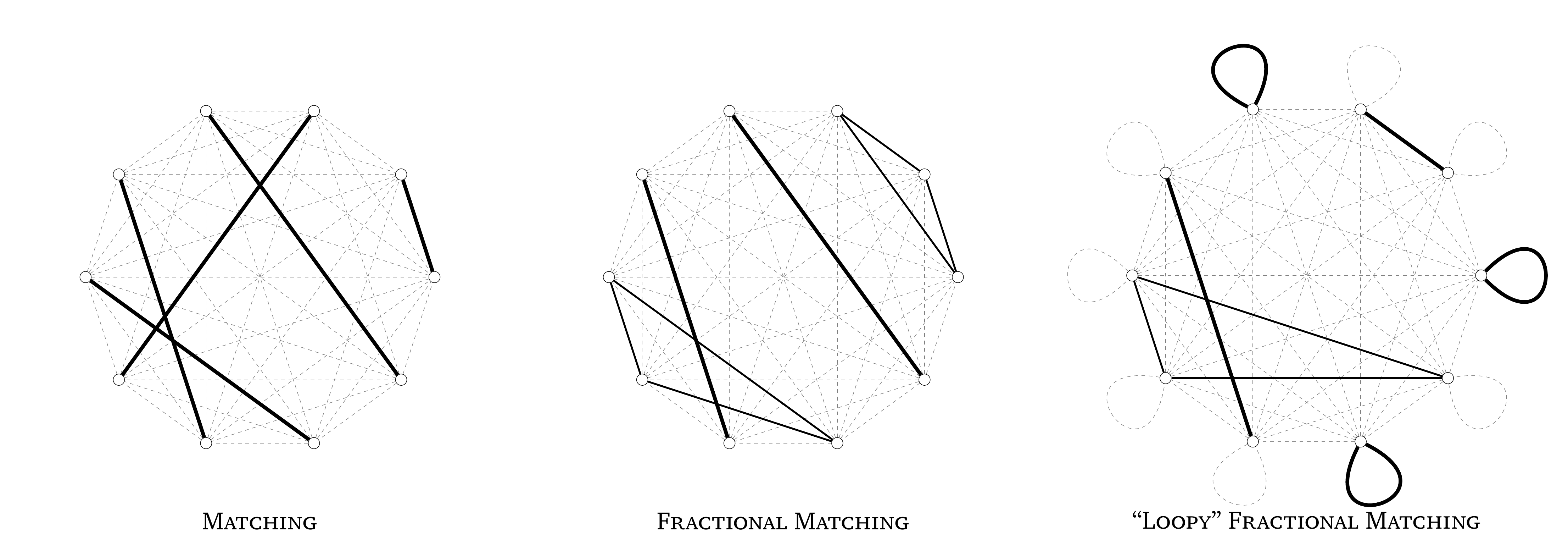}
 \caption{\textit{On the left}, complete graph $\mathcal K_{10}$ with a usual matching on it: the matching cost in this case is simply the sum of the weights $w_e$ of the matching edges $e\in\mathcal M$ (thick edges). \textit{In the center}, the same graph with a fractional matching on it: in this case, cycles are allowed and the contribution to the matching cost of an edge of weight $w_e$ belonging to a cycle is $\sfrac{w_e}{2}$ (thin edges), $w_e$ otherwise (thick edges). \textit{On the right}, the graph $\hat{\mathcal{K}}_{10}$ obtained allowing loops on each vertex of $\mathcal K_{10}$, with a fractional matching on it. From the left to the right, progressively more matching configurations are allowed. In particular, any feasible configuration for the usual matching problem is feasible for the fractional matching problem; moreover, any feasible configuration for the fractional matching problem is feasible for the loopy fractional matching problem.}\label{fig:match}
\end{figure}

Random-link matching problems have been the first class of optimization problems to be studied by statistical physics techniques, along with the random-link travelling salesman problem (\textsc{Rtsp}) \cite{Orland1985,Mezard1985}. The random-link minimum matching problem, simply called random matching problem (\textsc{Rmp}), is defined on the complete graph $\mathcal K_{2N}$, i.e., a graph having $2N$ vertices such that each of them is connected to all the others. In this case, the optimal matching is always perfect.  Each edge $e$ of the graph is associated to a random weight $w_e$, to be extracted, independently from all the others, according to a given probability density $\varrho$, that is assumed to be the same for all edges (see Fig.~\ref{fig:match}, left). In the \textsc{Rmp} we search for the set of ``edge occupations'' $\boldsymbol{\mathsf m}=\{m_e\}_e$ such that the cost (or energy) function
\begin{subequations}
\begin{equation}
E[\boldsymbol{\mathsf m},\boldsymbol{\mathsf w}]\coloneqq\sum_{e\in\mathcal E}m_e w_e,
\end{equation}
is minimized, with the constraints
\begin{equation}
 m_e\in\{0,1\}\quad\forall e\in\mathcal E,\quad\sum_{e\to v}m_e=1\quad\forall v\in\mathcal V.
\end{equation}
In the previous expression, the sum $\sum_{e\to v}$ runs over all edges having $v$ as an endpoint. Denoting by $\overline{\bullet}$ the average over all instances of the problem, it has been shown that, if $\lim_{w\to 0}\varrho(w)=1$ \cite{Mezard1985,Wastlund2008}
\begin{equation}
 \lim_{N\to+\infty}\overline{\min_{\boldsymbol{\mathsf m}} E[\boldsymbol{\mathsf m},\boldsymbol{\mathsf w}]}=\frac{\pi^2}{12}.
\label{asintE}
\end{equation}
\end{subequations}
The coefficient of the $\sfrac{1}{N}$ correction to this asymptotic result has been derived in Refs.~\cite{Mezard1987,Parisi2002,REMIDDI}, but its final expression is given in terms of an infinite series, whose terms have to be evaluated numerically, except for the first ones, for which an analytic expression is available.

In the bipartite version of the \textsc{Rmp}, called random assignment problem (\textsc{Rap}), the complete graph $\mathcal K_{2N}$ is replaced by a bipartite complete graph, $\mathcal K_{N,N}$, in which two sets of vertices, each one of cardinality $N$, are present, and each vertex of one set is connected to all vertices of the other set only. Both the asymptotic properties \cite{Mezard1985,Aldous2001} and the finite-size corrections to the average optimal cost (\textsc{aoc}) \cite{Parisi2002,Caracciolo2017} of the \textsc{Rap} have been extensively studied in the literature, and a closed formula is available for the \textsc{aoc} at any value of $N$ in the case of exponentially distributed weights \cite{Parisi1998,Nair2003,Linusson2004}, namely
\begin{equation}
 \overline{\min_{\boldsymbol{\mathsf m}} E[\boldsymbol{\mathsf m},\boldsymbol{\mathsf w}]}=\sum_{n=1}^{N}\frac{1}{n^2}=\frac{\pi^2}{6}-\frac{1}{N}+o\left(\frac{1}{N}\right).\end{equation}

Both the \textsc{Rmp} and the \textsc{Rap} play the role of mean-field versions of matching problems embedded in the Euclidean space, called random Euclidean matching problems, in which the vertices of the considered graphs correspond to random points in an Euclidean domain and the weight associated to an edge is a function of the Euclidean distance of the endpoints. This fact introduces correlation amongst the edge weights that has been treated as a correction to the purely random case \cite{Mezard1988,Lucibello2017}, and some analytical results have been obtained for the \textsc{aoc} or its finite-size corrections, using a direct field-theoretical approach \cite{Caracciolo2014,Caracciolo2015a,Caracciolo2015} or, in one dimension, purely probabilistic arguments \cite{Caracciolo2017b}. 
Analytical results regarding sparse instances of the matching problem \cite{Zhou2003a,Zdeborova2006,Bayati2006} have been obtained through the statistical physics cavity method \cite{mezard2009information}. 

In the present paper, we will apply the replica formalism to a different type of matching problem, namely the random fractional matching problem (\textsc{Rfmp}), the linear relaxation of the matching problem. We start considering a weighted complete graph $\mathcal K_{2N}$. The weights $\boldsymbol{\mathsf w}\coloneqq\{w_e\}_e$ associated to the edges $\{e\}_e\eqqcolon\mathcal E$ are non-negative independent random variables, identically distributed according to a probability density $\varrho(w)$, exactly as in the \textsc{Rmp} case. In the \textsc{Rfmp} we search for the set of quantities $\boldsymbol{\mathsf m}\coloneqq\{m_e\}_e$ that minimize the cost
\begin{subequations}
\begin{equation}
E[\boldsymbol{\mathsf m},\boldsymbol{\mathsf w}]\coloneqq\sum_{e\in\mathcal E}m_e w_e,
\end{equation}
with the additional constraints
\begin{equation}\label{costraintsm}
 m_e\in[0,1]\quad\forall e\in\mathcal E,\quad\sum_{e\to v}m_e=1\quad\forall v\in\mathcal V.
\end{equation}
\end{subequations}
It is easy to show for general graphs that the problem has semi-integer solutions, i.e. optimal configurations $\mathcal M$  with $m_e\in\{0,\sfrac{1}{2},1\}$. $\mathcal M$ contains only (odd) cycles and edges that do not share their endpoints  (see Fig.~\ref{fig:match}, center). It is expected that the \textsc{aoc} of the \textsc{Rfmp} is less than or equal to the \textsc{aoc} of the \textsc{Rmp} obtained with the same weight probability density $\varrho$, due to the fact that all matching configurations feasible for the \textsc{Rmp}, are also feasible for the \textsc{Rfmp}.

If defined on bipartite graphs, the relaxed problem above has integer solutions only, due to the absence of odd cycles, and therefore it is equivalent to the standard \textsc{Rap}. Noticeably, the Belief Propagation algorithm, deeply related to the cavity method, is able to recover the optimal solution of an assignment problem when the solution is unique, and its computational complexity is on par with that one of the best alternative solvers \cite{Bayati2008}. The statistical physics of linear or convex relaxations of discrete optimization problems has been actively investigated in recent years \cite{Takabe2016, Javanmard2016}.

In an interesting variation of the \textsc{Rfmp}, that we will call ``loopy \textsc{Rfmp}'', an additional non-negative weight $w_{v}$ is associated to each vertex $v\in\mathcal V$ of the graph. Each weight $w_v$ is a random variable extracted independently from all other weights in the problem with the same distribution of the edge weights $\varrho(w)$. The loopy \textsc{Rfmp} corresponds therefore to a \textsc{Rfmp} defined on a graph $\hat{\mathcal K}_{2N}$ obtained allowing self-loops on $\mathcal K_{2N}$ (see Fig.~\ref{fig:match}, right). The cost is defined as
\begin{subequations}
\begin{equation}
E[\boldsymbol{\mathsf m},\boldsymbol{\mathsf w}]\coloneqq\sum_{e\in\mathcal E}m_e w_e+2\sum_{v\in\mathcal V}m_v w_v,
\end{equation}
with the constraints
\begin{equation}\label{costraintsml}
 m_e\in[0,1]\quad\forall e\in\mathcal E,\quad m_v\in[0,1]\quad\forall v\in\mathcal V,\quad\sum_{e\to v}m_e+2m_v=1\quad\forall v\in\mathcal V.
\end{equation}
W{\"a}stlund proved that, in the loopy \textsc{Rfmp}, in the optimal configuration $m_{e}\in\{0,\sfrac{1}{2},1\}$ $\forall e$ and $m_v\in\{0,\sfrac{1}{2}\}$ $\forall v$ \cite{Wastlund2010}. Remarkably, he also obtained the expression for the \textsc{aoc} for any $N$ assuming $\varrho(w)=\e^{-w}\theta(w)$, namely
\begin{equation}\label{wastformula}
 \overline{\min_{\mathsf m}E[\boldsymbol{\mathsf m},\boldsymbol{\mathsf w}]}=\sum_{n=1}^{2N}\frac{(-1)^{n-1}}{n^2}=\frac{\pi^2}{12}-\frac{1}{8N^2}+o\left(\frac{1}{N^2}\right).
\end{equation}
\end{subequations}
Again, since the loopy \textsc{Rfmp} is a relaxed version of the \textsc{Rfmp}, that in turn is a relaxed version of the \textsc{Rmp}, it follows from Eq.~\eqref{wastformula} that all three \textsc{Rcop}s have the same asymptotic \textsc{aoc}, equal to $\sfrac{\pi^2}{12}$. The presence of cycles in the \textsc{Rfmp} and cycles and loops in the loopy \textsc{Rfmp} does not affect, therefore, the value of the \textsc{aoc}, but the finite-size corrections only.

In the present paper, we will study, using the replica approach, both the \textsc{Rfmp} and the loopy \textsc{Rfmp}, and their finite-size corrections. We shall consider weight probability densities $\varrho(w)$ with non-negative support and such that $\varrho(w)=1-\mu w+o(w)$ for $w\to 0^+$.  In particular, if the weights are exponentially distributed we have $\mu=1$, whereas $\mu=0$ if they are uniformly distributed. We will show, in particular, that the \textsc{Rmp} saddle-point solution naturally appears as asymptotic solution of the problem, alongside with another saddle-point solution of higher cost corresponding to the solution to the \textsc{Rtsp}. Moreover, we will evaluate the finite-size corrections to the \textsc{aoc} on the matching saddle-point, obtaining as the main result of the paper the closed formula
\begin{equation}\label{costofinaleintro}
E_{\varepsilon}(\mu)=\frac{\zeta(2)}{2}+\frac{1}{2N}\left[(\mu-1)\zeta(3)+\frac{1-\varepsilon}{4}\zeta(2)\right]+o\left(\frac{1}{N}\right).
\end{equation}
Here $\zeta(z)$ is the Riemann zeta function and, since $\zeta(2)=\sfrac{\pi^2}{6}$, we obtain the result of Eq.~\eqref{asintE} at the leading order. The parameter $\varepsilon$ takes value $+1$ if self-loops are allowed in the model, $-1$ otherwise, i.e., when $m_{ii}$ is not present. The analytic predictions from the replica calculation are supported by the numerical simulations in the last section of the paper.

\section{Replica calculation}
In the spirit of the seminal works of Orland \cite{Orland1985} and M\'ezard and Parisi \cite{Mezard1985}, let us first write down the partition function for the \textsc{Rfmp}, both in its usual and in its loopy version. As anticipated, we will consider a random-link formulation of the problem on the complete graph with $2N$ vertices $\mV=\{i\}_{i=1,\dots,2N}$, where the weights $\{w_{ij}\}_{ij}$ are non-negative independent random variables identically distributed with distribution $\varrho(w)$. As anticipated, in the following, we will consider a particular class of probability densities, i.e., we will assume that $\varrho(w)=1-\mu w+o(w)$ for $w\to 0^+$. The uniform distribution on the unit interval, $\varrho(w)=\theta(w)\theta(1-w)$, and the exponential distribution on the positive real axis, $\varrho(w)=\e^{-w}\theta(w)$, belong to this class, with $\mu=0$ and $\mu=1$ respectively. Due to the fact that all distributions in this class have the same limit for $w\to 0^+$, i.e., $\lim_{w\to 0^+}\varrho(w)=1$, we expect that the asymptotic \textsc{aoc} will be the same for all of them, as it happens in the \textsc{Rmp}. Indeed, in the general framework of the analysis of the \textsc{Rmp} performed in Ref.~\cite{Mezard1986,Parisi2002,Caracciolo2017} it is easily seen that, whereas the asymptotic cost only depends on the behavior of the first term in the Maclaurin expansion of $\varrho(w)\simeq\varrho_0 w^r$, the $O(\sfrac{1}{N})$ finite-size corrections depend on the expansion up to, at least, the second term \cite{Parisi2002}, and the power $r$ also affects the scaling of the corrections themselves \cite{Caracciolo2017}. 

Let us start observing that, in the \textsc{Rfmp}, the occupation number $m_{ij}=m_{ji}$ of the edge $(i,j)$ between the node $i$ and the node $j$ can assume the values
\begin{equation}
 m_{ij}\in\{0,1,2\},\quad \text{with the constraint } \sum_{j=1}^{2N}m_{ij}+\varepsilon m_{ii}=2,\quad 1\leq i\leq 2N,
\end{equation}
where $\varepsilon=+1$ if loops are allowed, $\varepsilon=-1$ otherwise. The parameter $\varepsilon$, therefore, allows us to switch between the two variations of the model described in the Introduction. The cost of a given matching configuration is
\begin{equation}\label{costoE}
 E_{\varepsilon}[\boldsymbol{\mathsf m},\boldsymbol{\mathsf w}]=\frac{1}{2}\sum_{i\leq j}\left(1+\varepsilon\delta_{ij}\right)m_{ij}w_{ij}.
\end{equation}
For calculation convenience we consider, for each edge $e$, $m_{e}\in\{0,1,2\}$ and not $m_{e}\in\{0,\sfrac{1}{2},1\}$ as in the Introduction. This fact does not affect the results apart from the necessary rescaling of the cost that indeed we have introduced in Eq.~\eqref{costoE}. The partition function for a given instance of the problem can be written as
\begin{equation}
 Z_\varepsilon(\upbeta)\coloneqq \sum_{m_{ij}\in\{0,1,2\}}\left[\prod_{i=1}^{2N}\mathbb I\left(\sum_{j=1}^{2N}m_{ij}+\varepsilon m_{ii}=2\right)\right]\e^{-2\upbeta N E_{\varepsilon}[\boldsymbol{\mathsf m},\boldsymbol{\mathsf w}]},
\end{equation}
where $\mathbb I(\bullet)$ is an indicator function that is equal to one when the condition in the brackets is satisfied, and zero otherwise. The \textsc{aoc} is recovered as
\begin{equation}
 E_{\varepsilon}(\mu)\coloneqq\overline{\min_{\boldsymbol{\mathsf m}}E_{\varepsilon}[\boldsymbol{\mathsf m},\boldsymbol{\mathsf w}]}=-\lim_{\upbeta\to+\infty}\frac{1}{2N}\frac{\partial\overline{\ln Z(\upbeta)}}{\partial\upbeta}.
\end{equation}
Note that we have made explicit the dependence of the \textsc{aoc} on the value of $\mu=-\left.\partial_w\varrho\right|_{w=0}$. From the results in Ref.~\cite{Wastlund2010}, we know that $E_{\varepsilon}(\mu)=O(1)$, i.e., the \textsc{aoc} is not extensive, and the cost density scales as $O\left(\sfrac{1}{N}\right)$: this is indeed expected, due to the fact that the shortest link amongst $N$ scales as $\sfrac{1}{N}$. We have therefore rescaled $\upbeta$ in the exponent of the partition function accordingly, in such a way that a finite thermodynamical limit at fixed $\upbeta$ can be obtained, and an extensive functional $2NE_{\varepsilon}(\mu)=O(N)$ appears in the exponent in the low-temperature regime. 

To average over the disorder, we use the following integral representation of the Kronecker delta,
\begin{equation}
 \delta_{a,0}=\int_{0}^{2\pi}\e^{i\lambda a}\frac{\dd\lambda}{2\pi},
\end{equation}
and we apply, as usual, the replica trick. The average replicated partition function for the fractional matching problem can be written as
\begin{equation}\label{Zpm}
 \overline{Z^n_\varepsilon}
 =\left[\prod_{a=1}^n\prod_{i=1}^{2N}\int_0^{2\pi}\frac{\e^{-2i\lambda_i^a}\dd\lambda_i^a}{2\pi}\right]\prod_{i<j}\left(1+\frac{T_{ij}}{N}\right)\prod_{i=1}^{2N}\left(1+\frac{\varepsilon+1}{2}\frac{R_{i}}{N}\right).
\end{equation}
In Eq.~\eqref{Zpm} we have introduced the quantities
\begin{subequations}\label{TR}
\begin{equation}
\begin{split}
 1+\frac{T_{ij}}{N}&\coloneqq \overline{\prod_{a=1}^n\left[1+\exp\left(i\lambda_i^a+i\lambda_j^a-\upbeta N  w_{ij}\right)+\exp\left(2i\lambda_i^a+2i\lambda_j^a-2\upbeta N w_{ij}\right)\right]}\\
 &=1+\frac{1}{N}\sum_{\mathclap{\substack{\alpha\cap\beta=\emptyset\\\alpha\cup\beta\neq\emptyset}}}\hat g_{|\alpha|+2|\beta|}\exp\left[i\sum_{a\in\alpha}\left(\lambda_i^a\!+\!\lambda_j^a\right)\!+\!2i\sum_{b\in\beta}\left(\lambda_i^b\!+\!\lambda_j^b\right)\right],
 \end{split}
\end{equation}
and, in the presence of loops, the on-site contribution
\begin{equation}
\begin{split}
 1+\frac{R_{i}}{N}&\coloneqq \overline{\prod_{a=1}^n\left[1+\exp\left(2i\lambda_i^a-2\upbeta N  w_{ii}\right)\right]}\\
 &=1+\frac{1}{N}\sum_{\mathclap{\alpha\neq\emptyset}}\hat g_{2|\alpha|}\exp\left[2i\sum_{a\in\alpha}\lambda_i^a\!\right].
 \end{split}
\end{equation}
\end{subequations}
In Eqs.~\eqref{TR} the sums run over the elements of $\mathcal P([n])$, set of subsets of $[n]\coloneqq\{1,\dots,n\}$, and we have denoted the cardinality of $\alpha\in\mathcal P([n])$ by $|\alpha|$. 
We have also introduced the quantity
\begin{equation}\label{gp}
\begin{split}
 \hat g_p&\coloneqq N\int_0^{+\infty}\e^{-\upbeta p N w}\varrho(w)\dd w\\
 &=g_p\left[1-\mu\frac{g_p}{N}+o\left(\frac{1}{N}\right)\right],\quad\text{where}\ g_p\coloneqq\frac{1}{\upbeta p}.
 \end{split}
\end{equation}
Using the previous equation, and in order to evaluate the \textsc{aoc} and its first finite-size correction, we use the fact that 
\begin{multline}
 2R_i=2\sum_{\mathclap{\alpha\neq\emptyset}}\hat g_{2|\alpha|}\exp\left[2i\sum_{a\in\alpha}\lambda_i^a\!\right]\\
 =\sum_{\mathclap{\alpha\neq\emptyset}}\left[g_{|\alpha|}-\frac{2g_{2|\alpha|}^2}{N}+o\left(\frac{1}{N}\right)\right]\exp\left[2i\sum_{a\in\alpha}\lambda_i^a\!\right]\sim T_{ii}+O\left(\frac{1}{N}\right),
\end{multline}
since $2g_{2p}=g_{p}$. In the last step we have used the fact that
\begin{equation}
T_{ii}=\sum_{\mathclap{\substack{\alpha\cap\beta=\emptyset\\\alpha\cup\beta\neq\emptyset}}}\hat g_{|\alpha|+2|\beta|}\exp\left[2i\sum_{a\in\alpha}\lambda_i^a\!+\!4i\sum_{b\in\beta}\lambda_i^b\right] 
\end{equation}
gives zero contribution, unless $\beta=\emptyset$, due to the overall constraint imposed by the integration on $\{\lambda_i^a\}$. Neglecting $O(\sfrac{1}{N})$ terms in the exponent (i.e., $O(\sfrac{1}{N^2})$ to the cost), we can write the partition function as
\begin{equation}
 \overline{Z^n_\varepsilon}
 =\left[\prod_{a=1}^n\prod_{i=1}^{2N}\int_0^{2\pi}\frac{\e^{-2i\lambda_i^a}\dd\lambda_i^a}{2\pi}\right]
 \exp\left[\frac{1}{2N}\sum_{i,j}\left(T_{ij}-\frac{T_{ij}^2}{2N}\right)+\frac{\varepsilon-1}{4N}\sum_{i=1}^{2N}T_{ii}+O\left(\frac{1}{N}\right)\right].
\end{equation}
We introduce now the placeholders
\begin{equation}
 q_{\alpha,\beta}\coloneqq \mathbb I(\alpha\cap\beta=\emptyset) \sum_{i=1}^{2N}\exp\left(i\sum_{a\in\alpha}\lambda_i^a+2i\sum_{b\in\beta}\lambda_i^b\right),
\end{equation}
that allow us to write
\begin{subequations}\label{TT2}
\begin{align}
 \sum_{i,j}T_{ij}+\frac{\varepsilon-1}{2}\sum_i T_{ii}&
 =\sum_{\mathclap{\substack{\alpha\cap\beta=\emptyset\\\alpha\cup\beta\neq\emptyset}}}\hat g_{|\alpha|+2|\beta|}
 q_{\alpha,\beta}^2+\frac{\varepsilon-1}{2}\sum_{\alpha\neq\emptyset}\hat g_{|\alpha|}q_{\emptyset,\alpha},\\
 \sum_{i,j}T_{ij}^2&=\sideset{}{'}\sum_{\mathclap{\alpha,\beta|\hat\alpha,\hat\beta}}\hat g_{|\alpha|+2|\beta|}\hat g_{|\hat \alpha|+2|\hat \beta|} q_{\alpha\triangle \hat \alpha,\beta\cup\hat\beta\cup(\alpha\cap\hat\alpha)}^2.\label{dst}
\end{align}
\end{subequations}
In the previous equations $\alpha\triangle\beta\coloneqq (\alpha\setminus\beta)\cup(\beta\setminus\alpha)$ is the symmetric difference of the sets $\alpha$ and $\beta$, we have denoted by 
\begin{equation}
 \sideset{}{'}\sum_{\alpha,\beta|\hat\alpha,\hat\beta}=\sum_{\mathclap{\substack{\alpha\cap\beta=\emptyset\\\alpha\cup\beta\neq\emptyset}}}\quad\sum_{\mathclap{\substack{\hat\alpha\cap\hat \beta=\emptyset\\\hat\alpha\cup\hat \beta\neq\emptyset}}}\mathbb I(\beta\cap\hat\beta=\emptyset)\mathbb I((\alpha\cup\hat\alpha)\cap(\beta\cup\hat\beta)=\emptyset).
\end{equation}
Denoting by
\begin{equation}
 \varphi_{\alpha,\beta}\coloneqq\sideset{}{'}\sum_{\sigma,\rho|\hat\sigma,\hat\rho}\hat g_{|\rho|+2|\sigma|}\hat g_{|\hat \rho|+2|\hat \sigma|}\mathbb I\left(\rho\triangle\hat\rho=\alpha\right)\mathbb I\left(\sigma\cup\hat\sigma\cup(\rho\cap\hat\rho)=\beta\right),
\end{equation}
and using a Hubbard--Stratonovich transformation in the form
\begin{multline}
 \exp\left(\frac{\hat g_{|\alpha|+2|\beta|}
 -\sfrac{1}{2N}\,\varphi_{\alpha,\beta}}{2N}q_{\alpha,\beta}^2\right)\\
 =\sqrt{\frac{N}{2\pi \left(\hat g_{|\alpha|+2|\beta|}\!-\!\sfrac{\varphi_{\alpha,\beta}}{2N}\right)}}\int_{-\infty}^{+\infty}\exp\left(-\frac{NQ_{\alpha,\beta}^2}{2\hat g_{|\alpha|+2|\beta|}\!-\!\sfrac{\varphi_{\alpha,\beta}}{N}}
 +Q_{\alpha,\beta}q_{\alpha,\beta}\right)\!\dd Q_{\alpha,\beta}\\
 \simeq \sqrt{\frac{N}{2\pi \hat g_{|\alpha|+2|\beta|}}}\int_{-\infty}^{+\infty}\exp\left(-\frac{NQ_{\alpha,\beta}^2}{2\hat g_{|\alpha|+2|\beta|}}
 -\frac{\varphi_{\alpha,\beta}Q^2_{\alpha,\beta}}{4\hat g^2_{|\alpha|+2|\beta|}}+Q_{\alpha,\beta}q_{\alpha,\beta}+O\left(\frac{1}{N}\right)\right)\dd Q_{\alpha,\beta}
\end{multline}
we can finally introduce the order parameters $Q_{\alpha,\beta}$ as follows
\begin{multline}\label{quisopra}
\left[\prod_{a=1}^n\prod_{i=1}^{2N}\int_0^{2\pi}\frac{\e^{-2i\lambda_i^a}\dd\lambda_i^a}{2\pi}\right]\exp\left[\frac{1}{2N}\sum_{i,j}\left(T_{ij}-\frac{T_{ij}^2}{2N}\right)+\frac{\varepsilon-1}{4N}\sum_{i=1}^{2N}T_{ii}\right]\\
\begin{split}
\simeq&\left[\prod_{\substack{\alpha\cap\beta=\emptyset\\\alpha\cup\beta\neq\emptyset}}\int\dd Q_{\alpha,\beta}\sqrt{\frac{N}{2\pi \hat g_{|\alpha|+2|\beta|}}}\right]\times\\
&\times\exp\left[-N\sum_{\mathclap{\substack{\alpha\cap\beta=\emptyset\\\alpha\cup\beta\neq\emptyset}}}\ \frac{Q_{\alpha,\beta}^2}{2\hat g_{|\alpha|\!+\!2|\beta|}}\!+\!2N\ln z[\bQ]\!
-\sideset{}{'}\sum_{\mathclap{\alpha,\beta|\hat\alpha,\hat\beta}}\frac{g_{|\alpha|\!+\!2|\beta|}g_{|\hat\alpha|\!+\!2|\hat\beta|}}{4 g^2_{|\alpha|\!+\!|\hat\alpha|\!+\!2|\beta|\!+\!2|\hat\beta|}}Q_{\alpha\triangle\hat\alpha,\beta\cup\hat\beta\cup(\alpha\cap\hat\alpha)}^2
 \right].
 \end{split}
\end{multline}
The expression of $\ln z[\bQ]$, in which we have exponentiated the integration on $\{\lambda_i^a\}$, is given in Eq.~\eqref{zone} in \ref{app:zonesite}. Eq.~\eqref{quisopra} generalizes the equivalent expression for the partition function obtained for the \textsc{Rmp} in Refs.~\cite{Mezard1985,Mezard1987,Parisi2002}, that is recovered imposing that $Q_{\alpha,\beta}\equiv Q_{\beta}\delta_{|\alpha|,0}$.

Observe that in Eq.~\eqref{quisopra} $\hat g_p$ appears, a quantity defined in Eq.~\eqref{gp}. We have that $\lim_{N\to +\infty}\hat g_p=g_p$. However, expanding for large $N$ the quantities $\hat g_p$ in Eq.~\eqref{quisopra}, new $\sfrac{1}{N}$ finite-size corrections to the cost will appear\footnote{In the case of a generic distribution $\varrho(w)=w^r[\varrho_0+\varrho_1 w+o(w)]$ these corrections will scale as  $\sfrac{1}{N^{\frac{k}{r+1}}}$ with $k\in\N$ and they will be \textsl{dominant} respect to all other corrections, that are $O(\sfrac{1}{N})$, for all values $k$ such that $k< r+1$, and of the same order for $k=r+1$ if $r+1$ is a natural number. In our case, only the $k=1$ term appears, because $r=0$, without any anomalous contribution. For further details on this point, see Ref.~\cite{Caracciolo2017}.}. In \ref{app:zonesite} we show that the replicated action can be finally written as
\begin{subequations}\label{S-dS}
\begin{multline}\label{ints}
 \overline{Z^n}
 \simeq\left[\prod_{\substack{\alpha\cap\beta=\emptyset\\\alpha\cup\beta\neq\emptyset}}\int\dd Q_{\alpha,\beta}\sqrt{\frac{N}{2\pi \hat g_{|\alpha|+2|\beta|}}}\right]\e^{-N\mathcal S[\bQ]}\\
 \mathcal S[\bQ]\coloneqq S[\bQ]+\Delta S^\text{T}[\bQ]+\Delta S^{\varrho}[\bQ]+o\left(\sfrac{1}{N}\right).
\end{multline}
The three contributions appearing in the previous expression are
\begin{align}
 S[\bQ]&\coloneqq \sum_{\substack{\alpha\cap\beta=\emptyset\\\alpha\cup\beta\neq\emptyset}}\frac{Q_{\alpha,\beta}^2}{2g_{|\alpha|+2|\beta|}}-2\ln z_0[\bQ],\label{s0}\\
 \Delta S^\text{T}[\bQ]&=\frac{1}{N}\sideset{}{'}\sum_{\mathclap{\alpha,\beta|\hat\alpha,\hat\beta}}\frac{g_{|\alpha|+2|\beta|}g_{|\hat\alpha|+2|\hat \beta|}}{4 g^2_{|\alpha\triangle \hat\alpha|+2|\beta|+2|\hat\beta|+2|\alpha\cap\hat\alpha|}}Q_{\alpha\triangle\hat \alpha,\beta\cup\hat\beta\cup(\alpha\cap\hat\alpha)}^2-\frac{\varepsilon-1}{2N}\sum_{\alpha\neq\emptyset}g_{|\alpha|}\frac{\partial \ln z_0[\bQ]}{\partial Q_{0,\alpha}},\\
 \Delta S^{\varrho}[\bQ]&\coloneqq \frac{\mu}{2\upbeta N}\sum_{\substack{\alpha\cap\beta=\emptyset\\\alpha\cup\beta\neq\emptyset}}Q_{\alpha,\beta}^2,\end{align}
where
\begin{equation}\label{z0}
 z_0[\bQ]\coloneqq\lim_{\mathclap{N\to\infty}} z[\bQ]=  \left[\prod_{a=1}^n\int_{0}^{2\pi}\frac{\e^{-2i\lambda^a}\dd\lambda^a}{2\pi}\right] \exp\left[\sum_{{\substack{\alpha\cap\beta=\emptyset\\\alpha\cup\beta\neq\emptyset}}}Q_{\alpha,\beta}\exp\left(i\sum_{a\in\alpha}\lambda^a+2i\sum_{b\in\beta}\lambda^b\right)\right]
\end{equation}\end{subequations}
is the (leading) one-site partition function. It follows that $S$ is the leading term in $\mathcal S$. The $\Delta S^\text{T}$ term contains the finite-size correction due to the re-exponentiation and to the one-site partition function $z$. Finally, the $\Delta S^{\varrho}$ term contains an additional contribution due to the finite-size corrections to $g_p$ appearing in Eq.~\eqref{gp}. Observe, once again, that this contribution is absent in the case of flat distribution.

The integral in Eq.~\eqref{ints} can now be evaluated using the saddle-point method. The saddle-point equation for $Q_{\alpha,\beta}$ is
\begin{equation}\label{saddlepointQ}
 \frac{Q_{\alpha,\beta}}{g_{|\alpha|+2|\beta|}}=2\frac{\partial \ln z_0[\bQ]}{\partial Q_{\alpha,\beta}}=2\left\langle\exp\left(i\sum_{a\in\alpha}\lambda^a+2i\sum_{b\in\beta}\lambda^b\right)\right\rangle_{z_0},
\end{equation}
where $\langle\bullet\rangle_{z_0}$ is the average performed respect to the one-site partition function $z_0$. Denoting by $\bQ^\text{sp}$ the solution of Eq.~\eqref{saddlepointQ}, the replicated partition function becomes
\begin{equation}\label{sfinale}
 \overline{Z^n}\simeq \exp\left(-N\mathcal S[\bQ^\text{sp}]-\frac{\ln\det\boldsymbol{\mathsf \Omega}[\bQ^\text{sp}]}{2}\right),
\end{equation}
where $\boldsymbol{\mathsf \Omega}$ is the Hessian matrix of $S[\bQ]$ evaluated on the saddle-point $\bQ^\text{sp}$, solution of Eq.~\eqref{saddlepointQ}, i.e.,
\begin{equation}\label{Omega}
 \Omega_{\alpha\beta,\hat\alpha\hat\beta}[\bQ^\text{sp}]\coloneqq \sqrt{g_{|\alpha|+2|\beta|}g_{|\hat\alpha|+2|\hat\beta|}}\left.\frac{\partial^2 S[\bQ]}{\partial Q_{\alpha,\beta}\partial Q_{\hat\alpha,\hat \beta}}\right|_{\bQ=\bQ^\text{sp}}
\end{equation}
The additional term $-\sfrac{1}{2}\ln\det\boldsymbol{\mathsf \Omega}[\bQ^\text{sp}]$ provides, in general, a nontrivial finite-size correction to the leading term \cite{Mezard1987,Parisi2002}.

\subsection{Replica symmetric ansatz and matching saddle-point}
To proceed further, let us assume that a replica-symmetric ansatz holds, i.e., we search for a solution of the saddle-point equation in the form
\begin{equation}
 Q_{\alpha,\beta}\equiv Q_{|\alpha|,|\beta|}.
\end{equation}
This is a common and successful assumption in the study of random combinatorial optimization problems \cite{Mezard1985,Mezard1986,Mezard1988} that greatly simplifies the calculation. The stability of a replica symmetric solution is however not obvious and, in general, a replica symmetry breaking might occur \cite{mezard2009information}. In the present paper, this assumption will be justified \textsl{a posteriori}, on the basis of the agreement between the analytical predictions and the numerical computation. Using the replica symmetric hypothesis, the leading term in Eq.~\eqref{s0} becomes
\begin{multline}
 S[\bQ]=\sum_{p+q\geq 1}\binom{n}{p\ q}\frac{Q_{p,q}^2}{2g_{p+2q}}-2\ln z_0[\bQ]\\
 \xrightarrow{n\to 0} n\sum_{\mathclap{p+q\geq 1}}\frac{(-1)^{p+q-1}}{p+q}\binom{p+q}{q}\frac{Q_{p,q}^2}{2g_{p+2q}}-2n\lim_{n\to 0}\frac{\ln z_0[\bQ]}{n}.\label{Snsmall}
\end{multline}
In the previous expression, and in the following, we will adopt the notation
\begin{equation}
 \binom{a}{b_1\cdots b_s}\coloneqq\frac{\Gamma(a+1)}{\Gamma\left(a+1-\sum_{i=1}^sb_i\right)\prod_{i=1}^s\Gamma(b_i+1)}.
\end{equation}
Even under the replica symmetric hypothesis, the evaluation of $\lim_{n\to 0}\frac{1}{n}\ln z_0[\bQ]$ remains nontrivial. However, in \ref{app:zonesite} we show that a special replica symmetric saddle-point solution exists, namely
\begin{equation}\label{matchingsp}
 Q_{p,q}^\text{sp}=\delta_{p,0}Q_{0,q}\equiv \delta_{p,0}Q_q,
\end{equation}
corresponding to the replica-symmetric saddle-point solution of the \textsc{Rmp}. This fact is not surprising: as anticipated in the Introduction, the \textsc{aoc} of the \textsc{Rfmp} coincides with the \textsc{aoc} of the \textsc{Rmp} in the $N\to+\infty$ limit, and, indeed, the evaluation of $z_0[\bQ]$ on the matching saddle-point can be performed exactly, and coincides with the one of the \textsc{Rmp} \cite{Mezard1985,Mezard1987,Parisi2002}
\begin{subequations}\label{zmatchingsp}
\begin{align}
 \label{zmatchingsp1}\ln z_0[\bQ^\text{sp}]&=n\int_{-\infty}^{+\infty}\left(\e^{-\e^x}-\e^{-G(x)}\right)\dd x,\\
 \label{zmatchingsp2}G(x)&\coloneqq\sum_{k=1}^\infty \frac{(-1)^{k-1}}{k!}Q_k\e^{xk}.
\end{align}
\end{subequations}
The saddle-point equations become
\begin{equation}
Q_{p,q}=\delta_{p,0}Q_q=\frac{\delta_{p,0}}{\upbeta}\int_{-\infty}^{+\infty}\frac{\e^{qy-G(y)}}{q!}\dd y,
\end{equation}
implying the self-consistent equation for $G$ given by
\begin{align}
 G(x)= \frac{1}{\upbeta}\int_{-\infty}^{+\infty}B(x+y)\e^{-G(y)}\dd y,\\
 B(x)\coloneqq \sum_{k=1}^\infty (-1)^{k-1} \frac{\e^{kx}}{\Gamma^2(k+1)}.
\end{align}
As expected, the leading contribution corresponds therefore to the \textsc{Rmp} free-energy,
\begin{multline}
 \frac{1}{n}S[\bQ^\text{sp}]=\upbeta\sum_{k=1}^\infty (-1)^{k-1} Q_{k}^2 -2\int_{-\infty}^{+\infty}\left(\e^{-y}-\e^{-G(y)}\right)\dd y\\
 =\int_{-\infty}^{+\infty}G(y)\e^{-G(y)}\dd y-2\int_{-\infty}^{+\infty}\left(\e^{-y}-\e^{-G(y)}\right)\dd y.
\end{multline}
In the $\upbeta\to+\infty$ limit we can introduce
\begin{equation}
 \hat G(x)\coloneqq G\left(\upbeta x\right).
\end{equation}
and we can use the fact that \cite{Mezard1985}
\begin{equation}
 \lim_{\upbeta\to+\infty} B(\upbeta x)=\theta(x).
\end{equation}
We can solve for $\hat G$ as
\begin{equation}\label{saddleG}
 \hat G(x)=\int_{-x}^{+\infty}\e^{-\hat G(y)}\dd y\Longrightarrow \hat G(x)=\ln(1+\e^x).
\end{equation}
We finally obtain that, in both the considered formulations, the \textsc{aoc} of the \textsc{Rfmp} is equal to the \textsc{aoc} of the \textsc{Rmp}, as anticipated in the Introduction,
\begin{equation}
 \lim_{N\to+\infty}E_\varepsilon(\mu)=\frac{1}{2}\int_{-\infty}^{+\infty}\hat G(l)\e^{-\hat G(l)}\dd l=\frac{\pi^2}{12}.
\end{equation}

\subsection{Finite-size corrections}
In Eq.~\eqref{sfinale} three contributions to the finite-size corrections appear, namely $\Delta S^\varrho$, $\Delta S^\text{T}$, and $\sfrac{1}{2}\ln\det\boldsymbol{\mathsf\Omega}$. The first contribution can be evaluated straightforwardly as
\begin{equation}
\begin{split}
 \Delta S^{\varrho}[\bQ^\text{sp}]&=\frac{\mu}{2N}\sum_{\beta\neq \emptyset} Q_{|\beta|}^2 =\frac{\mu}{2N}\sum_{q=1}^\infty\binom{n}{q} Q_{q}^2=\frac{n\mu}{2N}\sum_{q=1}^\infty\frac{(-1)^{q-1}}{q} Q_{q}^2+o(n)\\
 &=\frac{n\mu}{2N\upbeta}\int_{-\infty}^{+\infty}\dd y\int_0^{+\infty}\dd x\,G(y-x)\e^{-G(y)}+o(n).
 \end{split}
\end{equation}
In the $\upbeta\to+\infty$ limit we obtain
\begin{equation}
 \lim_{\upbeta\to+\infty}\lim_{n\to 0}\frac{\Delta S^{\varrho}[\bQ^\text{sp}]}{n\upbeta}=\frac{\mu}{2N}\int_{-\infty}^{+\infty}\dd y\int_0^{+\infty}\dd x\,\hat G(y-x)\e^{-\hat G(y)}=\frac{\mu}{N}\zeta(3).
\end{equation}
The last integration was performed in Ref.~\cite{Mezard1987}. The $\Delta S^\mathrm{T}$ contribution depends on $\varepsilon$, i.e., on the presence or not of loops. In \ref{app:dst} we show that
\begin{equation}
 \lim_{\upbeta\to+\infty}\lim_{n\to 0}\frac{\Delta S^\text{T}[\bQ^\text{sp}]}{n\upbeta}=-\frac{\varepsilon-1}{4N}\zeta(2)-\frac{1}{N}\zeta(3).
\end{equation}
Finally, the fluctuation contribution 
\begin{equation}
\frac{\Delta E^{\boldsymbol{\mathsf \Omega}}}{N}\coloneqq \lim_{\upbeta\to+\infty}\lim_{n\to 0}\frac{\ln\det\boldsymbol{\mathsf \Omega}[\bf Q^\text{sp}]}{4n\upbeta}.
\end{equation}
The expression of this fluctuation term is more involved than the corresponding one for the \textsc{Rmp}, that has been studied in Refs.~\cite{Mezard1987,Parisi2002}. As it happens in that case, an exact evaluation through the replica formalism is quite complicated (we give some details in \ref{app:determinante}). However, observe that $\Delta E^{\boldsymbol{\mathsf \Omega}}$ depends on neither $\varepsilon$ nor $\mu$. We can therefore avoid the complex, direct evaluation and extract its value from W{\"a}stlund's formula in Eq.~\eqref{wastformula} for the \textsc{aoc} in the loopy \textsc{Rfmp} with exponentially distributed weights, comparing our result in this specific case with W{\"a}stlund's one. In particular, Eq.~\eqref{wastformula} predicts no $\sfrac{1}{N}$ corrections in the loopy \textsc{Rfmp}, implying the simple result
\begin{equation}
\Delta E^{\boldsymbol{\mathsf \Omega}}=0.
\end{equation}
Moreover, in the spirit of the analysis in Ref.~\cite{DeAlmeida1978}, the fact that $\Delta E^{\boldsymbol{\mathsf \Omega}}$ is a finite and well-defined quantity also suggests that the Hessian $\boldsymbol{\mathsf \Omega}$ remains positive definite within the replica symmetric ansatz for $\upbeta\to+\infty$, and, therefore, that the replica symmetric solution remains stable.

Collecting all contributions, we can finally write down a general expression for the \textsc{aoc} for the \textsc{Rfmp}, and its finite-size corrections, as
\begin{equation}\label{asintoticafinale}
E_{\varepsilon}(\mu)=\frac{\zeta(2)}{2}+\frac{1}{2N}\left[(\mu-1)\zeta(3)+\frac{1-\varepsilon}{4}\zeta(2)\right]+o\left(\frac{1}{N}\right).
\end{equation}

\section{Numerical results}
The analytic results in Eq.~\eqref{asintoticafinale} have been verified numerically using the \texttt{LEMON} graph library \cite{lemon}. For each one of the considered models, the results have been obtained averaging over at least $3\cdot 10^6$ instances for each value of $N$.
\begin{figure}\label{fig:costo}
\subfloat[\textsc{Aoc} for the \textsc{Rmp}, the \textsc{Rfmp} and the loopy \textsc{Rfmp} in the case of uniform distribution for the weights ($\mu=0$). The smooth lines correspond to the fits obtained using the asymptotic theoretical predictions for both the asymptotic \textsc{aoc} and the $\sfrac{1}{N}$ corrections in the \textsc{Rfmp} and the loopy \textsc{Rfmp} in Eq.~\eqref{fitrfmp} (see last column of Table~\ref{tab:fit}). The black line corresponds to the fit obtained using Eq.~\eqref{fitrmp} for the \textsc{Rmp} data (see Table~\ref{tab:fit}).]{\includegraphics[width=0.48\textwidth]{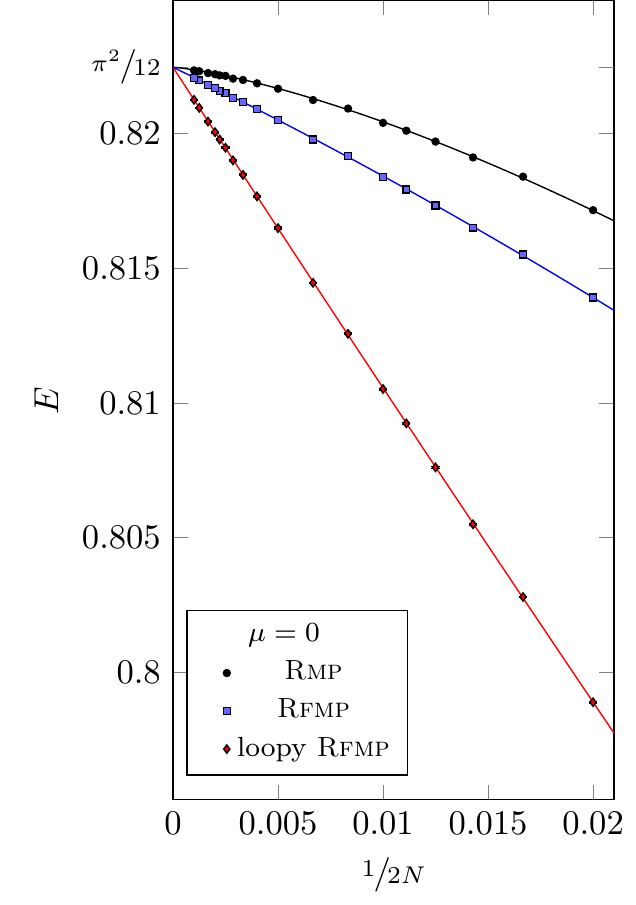}}\hfill
\subfloat[\textsc{Aoc} for the \textsc{Rmp}, the \textsc{Rfmp} and the loopy \textsc{Rfmp} in the case of exponential distribution for the weights ($\mu=1$). The blue line corresponds to the fit obtained using the asymptotic theoretical prediction for the \textsc{Rfmp} in Eq.~\eqref{fitrfmp} (see last column of Table~\ref{tab:fit}). The red line is W{\"a}stlund's formula in Eq.~\eqref{wastformula} for the loopy \textsc{Rfmp}, that is exact for all values of $N$. Finally, the black line corresponds to the fit obtained using Eq.~\eqref{fitrmp} for the \textsc{Rmp} data (see Table~\ref{tab:fit}). ]{\includegraphics[width=0.48\textwidth]{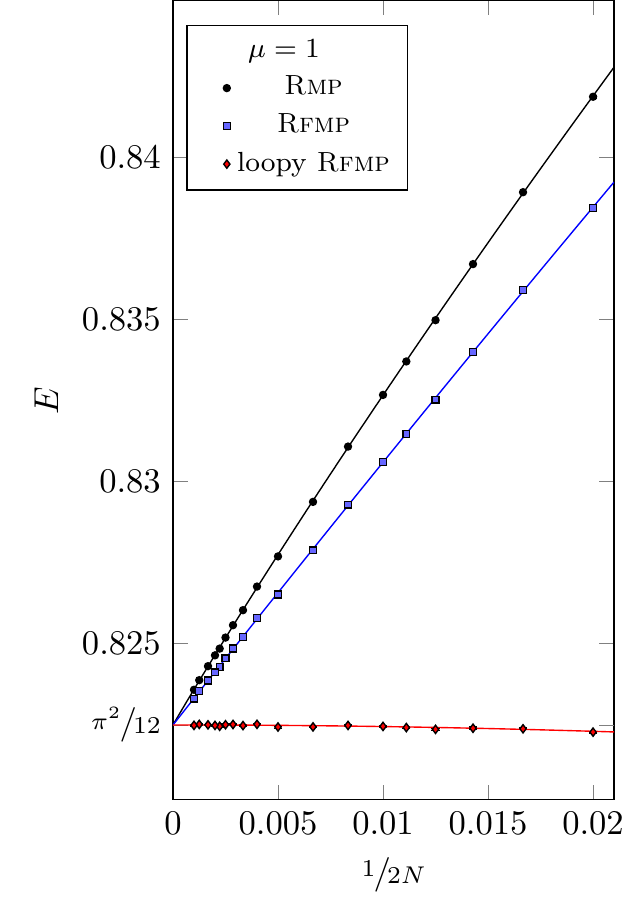}}
 \caption{\textsc{Aoc} for the \textsc{Rmp}, the \textsc{Rfmp} and the loopy \textsc{Rfmp} in the case of both uniform and exponential distribution for the weights. Error bars are represented but smaller than the markers.}
\end{figure}
In Fig.~\ref{fig:costo} we show our results, both for the case of uniform weight distribution ($\mu=0$) and the case of exponential weight distribution ($\mu=1$). In both cases it is evident that the \textsc{aoc} of the \textsc{Rmp} is greater than the corresponding \textsc{aoc} of the \textsc{Rfmp}, and similarly $E_{-1}(\mu)\geq E_{+1}(\mu)$, as expected. The asymptotic formula, Eq.~\eqref{asintoticafinale}, has been verified for all cases. We performed moreover a parametric fit of our data using the fitting function 
\begin{subequations}
\begin{equation}\label{fitrfmp}
 E=\frac{\pi^2}{12}+\frac{a}{2N}+\frac{b}{4N^2}+\frac{c}{8N^3}
\end{equation}
to verify our predictions for the \textsc{Rfmp} in two ways, i.e., either assumiming all parameters free, or fixing $a$ equal to our analytical prediction (when available) to improve our evaluation of $b$ and $c$. We also performed, for comparison, a similar fit for the \textsc{Rmp}, using the fitting function \begin{equation}\label{fitrmp}
 E=\frac{\pi^2}{12}+\frac{a}{2N}+\frac{\hat c}{(2N)^{\sfrac{3}{2}}}+\frac{b}{4N^2},
\end{equation}
\end{subequations}
to take into account the anomalous scaling of the corrections in this case \cite{Lucibello2017}. The results are summarized in Table~\ref{tab:fit}. It is remarkable that the $N^{-\sfrac{3}{2}}$ correction introduced in Eq.~\eqref{fitrfmp} for the \textsc{Rmp} cost, that is numerically present in agreement with the results in Ref.~\cite{Lucibello2017}, is absent in all considered variants of the \textsc{Rfmp}, as we numerically verified and as analytically predicted in Eq.~\eqref{wastformula} for the loopy \textsc{Rfmp} with exponential weight distribution.

\begin{table}\centering
\resizebox{\textwidth}{!}{  
\begin{tabular}{c|c|c|ccc|cc}
\toprule[1.5pt]
&\multirow{2}{*}{\textsc{Problem}}&Theoretical&\multicolumn{3}{c|}{$a$, $b$, $c$/$\hat c$ free}&\multicolumn{2}{c}{$b$, $c$ free}\\
&&$a$&$a$&$b$&$c$ or $\hat c$&$b$&$c$\\
    \midrule[1.5pt]
    \multirow{3}{*}{$\mu=0$} & \textsc{Rmp} & --&$-0.049(4)$&$0.03(9)$&$-1.54(4)$&--&-- \\
     & \textsc{Rfmp} $\varepsilon=+1$ & $-1.2020569\dots$&$-1.204(1)$&$1.26(5)$&$-1.0(4)$&$1.19(2)$&$-0.5(2)$ \\
     & \textsc{Rfmp} $\varepsilon=-1$ & $-0.3795898\dots$&$-0.381(1)$&$-2.43(5)$&$4.7(4)$&$-2.50(2)$&$5.2(2)$ \\
    \hline
    \multirow{3}{*}{$\mu=1$} & \textsc{Rmp} & --&$1.131(3)$&$-0.03(8)$&$-1.13(3)$&--&-- \\
     & \textsc{Rfmp} $\varepsilon=+1$ & $0$&$-0.001(1)$&$-0.48(5)$&$0.5(5)$&$-0.53(2)$&$0.9(2)$\\
     & \textsc{Rfmp} $\varepsilon=-1$ & $0.8224670\dots$&$0.821(1)$&$-1.15(6)$&$0.5(5)$&$-1.19(3)$&$0.9(3)$ \\
    \bottomrule[1.5pt]
  \end{tabular}}
\caption{Results of a fitting procedure of the \textsc{aoc} obtained numerically compared with the theoretical predictions. W{\"a}stlund's formula predicts $a=0$ and $b=-c=-\sfrac{1}{2}$ for the \textsc{aoc} of the \textsc{Rfmp} with $\mu=1$ and $\epsilon=+1$. \label{tab:fit}}
\end{table}

\begin{figure}\label{fig:diffcosto}
\subfloat[Difference $\delta E_\ell$ between the \textsc{aoc} obtained with loops and the \textsc{aoc} obtained without loops, both for $\mu=1$ (exponentially distributed weights) and for $\mu=0$ (uniformly distributed weights). The smooth line is the predicted asymptotic behavior given Eq.~\eqref{diffloops}.]{\includegraphics[width=0.48\textwidth]{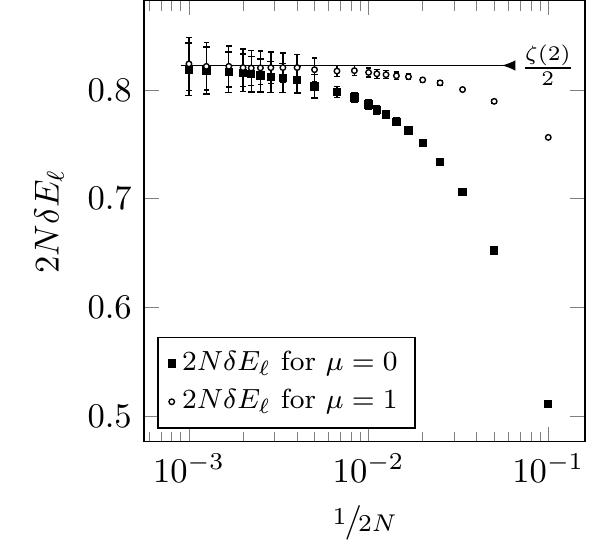}}
\hfill
\subfloat[Difference $\delta E_\varrho$ between the \textsc{aoc} obtained with exponentially distributed weights and the \textsc{aoc} obtained with uniformly distributed weights, both for $\epsilon=1$ (with loops) and $\varepsilon=-1$ (without loops). The smooth line is the predicted asymptotic behavior given Eq.~\eqref{diffrho}.]{\includegraphics[width=0.48\textwidth]{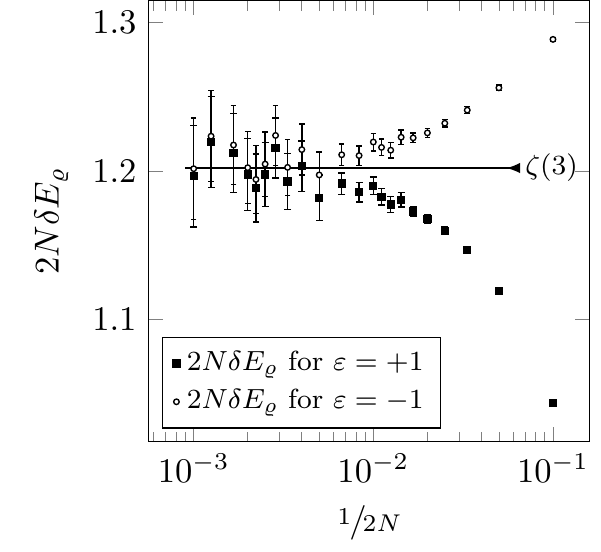}} 
\caption{Plots of the differences $\delta E_\ell$ and $\delta E_\varrho$ as functions of $N$. In both plots error bars are represented but smaller than the markers.}
\end{figure}

Eq.~\eqref{asintoticafinale} allows us to make predictions about differences of \textsc{aoc} for different types of models, due to the fact that we have isolated the different contributions depending on the presence of loops, or on the chosen distribution $\varrho$. For example, we expect that
\begin{subequations}\label{diffeq}
\begin{equation}\label{diffloops}
 \delta E_\ell\coloneqq E_{-1}(\mu)-E_{+1}(\mu)=\frac{\zeta(2)}{4N}+o\left(\frac{1}{N}\right).
\end{equation}
Similarly, we have that
\begin{equation}\label{diffrho}
 \delta E_\varrho\coloneqq E_{\varepsilon}(1)-E_{\varepsilon}(0)=\frac{\zeta(3)}{2N}+o\left(\frac{1}{N}\right). 
\end{equation}
\end{subequations}
Both the relations above have been verified numerically. Our results are shown in Fig.~\ref{fig:diffcosto} and they are in agreeement with Eqs.~\eqref{diffeq}.

\section{Conclusions}
The \textsc{Rfmp} on the complete graph $\mathcal K_{2N}$ generalizes the more famous \textsc{Rmp}, allowing cycles in the optimal solution. Here we have studied, using the replica formalism, its \textsc{aoc} and we have compared it with the \textsc{aoc} of the \textsc{Rmp}. We have considered the case of random weights on the graph edges distributed according to $\varrho(w)=1-\mu w+o(w)$, and we have evaluated the asymptotic expression of the \textsc{aoc} on the matching saddle-point, obtaining for the \textsc{Rfmp} the same asymptotic \textsc{aoc} of the \textsc{Rmp}. Remarkably enough, another saddle-point solution naturally appears in the calculation, corresponding to the \textsc{Rtsp} solution, whose average optimal cost is however higher. 

We have also explicitly obtained the finite-size corrections to the asymptotic \textsc{aoc} in two variations of the model, the standard \textsc{Rfmp} and the loopy \textsc{Rfmp}, in which each node can be matched to itself (loop). For the latter model, in particular, an explicit formula for the \textsc{aoc} had been found by W{\"a}stlund for any value of $N$ in the case of exponentially distributed weights \cite{Wastlund2010}. The two models have different, non-trivial finite-size corrections. 

By means of our approach, we have been able to separate the different contributions in the finite size corrections coming from different details of the problem, namely a first one due to the possibility of having loops, a second one due to the chosen weight distribution and a third one that is independent from the aforementioned characteristics. The first and the second contributions have been explicitly derived, whereas the third one has been obtained comparing W{\"a}stlund's result with ours in the corresponding specific case. We have been able to write down an explicit expression for the \textsc{aoc} for all the considered cases up to $o(\sfrac{1}{N})$ corrections, given in Eq.~\eqref{costofinaleintro}. We finally verified our results in all the considered cases comparing them with the values for the \textsc{aoc} obtained numerically. Going beyond the $\sfrac{1}{N}$ corrections, numerical results also suggest that the anomalous scaling $\sfrac{1}{N^{\sfrac{3}{2}}}$ in higher order corrections in the \textsc{Rmp}, discussed in Ref.~\cite{Lucibello2017} and verified in the present work, does not appear in the \textsc{Rfmp}.

Despite the fact that an explicit expression for the \textsc{aoc} has been found for the \textsc{Rfmp} in all the considered formulations of it, a number of open problems still remains. For example, a complete replica calculation of the contribution $\Delta E^{\boldsymbol{\mathsf \Omega}}$ deriving from $\ln\det\boldsymbol{\mathsf \Omega}$ is still missing. We have been able to estimate this contribution relying on W{\"a}stlund's results. However, an explicit replica calculation is still interesting for a series of reasons. One of them is that, as shown in \ref{app:determinante}, $\Delta E^{\boldsymbol{\mathsf \Omega}}$ appears to be the sum of two terms, the first one identical to a corresponding one appearing in the \textsc{Rmp} that is known to be nonzero, the second one that we expect, from W{\"a}stlund's formula, to be exactly opposite to the former one, in such a way that $\Delta E^{\boldsymbol{\mathsf \Omega}}=0$. The presence of this cancellation suggests that the contribution of cycles plays an important role in the anomalous scaling of higher corrections in the \textsc{Rmp}, and in its absence in the \textsc{Rfmp}. Remembering also that an explicit formula for the $\sfrac{1}{N}$ correction in the \textsc{Rmp} is still missing, further investigations in this direction are in order, to shed some light on the problem of finite-size corrections both in the \textsc{Rfmp} and in the \textsc{Rmp} even beyond the $O(\sfrac{1}{N})$ terms.

\section*{Acknowledgments}
The authors are grateful to Johan W{\"a}stlund for useful discussions and correspondence. GP and GS acknowledge the financial support of the Simons foundation (Grant No.~454949, Giorgio Parisi). The work presented in this article was supported by the project ``Meccanica statistica e complessit\`a'', a research grant funded by PRIN 2015 (Agreement no.~2015K7KK8L).

\appendix
\section{One-site partition function}\label{app:zonesite}
The evaluation of the one-site partition function in the \textsc{Rfmp} follows the same type of arguments adopted in the literature for the \textsc{Rmp} \cite{Mezard1985,Mezard1987,Parisi2002} and for the \textsc{Rtsp} \cite{Mezard1986}. The one-site partition function $z$ in the \textsc{Rfmp} is
\begin{multline}\label{zone}
 z[\bQ]\coloneqq\prod_{a=1}^n \left[\int_{0}^{2\pi}\frac{\e^{-2i\lambda^a}\dd\lambda^a}{2\pi}\right]\\\times \exp\left[\sum_{\substack{\alpha\cap\beta=\emptyset\\\alpha\cup\beta\neq\emptyset}}Q_{\alpha,\beta}\exp\left(i\sum_{a\in\alpha}\lambda^a+2i\sum_{b\in\beta}\lambda^b\right)+\frac{\varepsilon-1}{4N}\sum_{\alpha\neq\emptyset}g_{|\alpha|}\exp\left(2i\sum_{a\in\alpha}\lambda^a\right)
 \right]\\
 \equiv z_0[\bQ]+\frac{\varepsilon-1}{2N}\sum_{\alpha\neq\emptyset}g_{|\alpha|}\frac{\partial z_0[\bQ]}{\partial Q_{\emptyset,\alpha}}.
\end{multline}
In the previous equation, we have introduced $z_0[\bQ]$, that coincides with the expression given in Eq.~\eqref{z0}. It follows that, for $N\gg 1$, $\ln z[\bQ]$ provides a contribution both to the leading term $S[\bQ]$ and to the finite-size corrections $\Delta S^\text{T}[\bQ]$, namely, up to $o\left(\sfrac{1}{N}\right)$ terms,
\begin{equation}
 2\ln z[\bQ]=2\ln z_0[Q]-\frac{\varepsilon-1}{2N}\sum_{\alpha\neq\emptyset}\frac{g_{|\alpha|}}{z_0[\bQ]}\frac{\partial z_0[\bQ]}{\partial Q_{\emptyset,\alpha}},
\end{equation}
to be compared with the terms appearing in Eqs.~\eqref{S-dS}. The evaluation of $z_0[\bQ]$ is nontrivial in general. In the replica symmetric hypothesis, $z_0[\bQ]$ can be written as
 \begin{equation}
z_0[\bQ]=\prod_{a=1}^n \left[\int_{0}^{2\pi}\frac{\e^{-2i\lambda^a}\dd\lambda^a}{2\pi}\right] \exp\left[\sum_{{\substack{\alpha\cap\beta=\emptyset\\\alpha\cup\beta\neq\emptyset}}}Q_{|\alpha|,|\beta|}\exp\left(i\sum_{a\in\alpha}\lambda^a+2i\sum_{b\in\beta}\lambda^b\right)\right].
 \end{equation} 
Denoting now by $\xi_a\coloneqq \e^{i\lambda_a}$, observe now that
\begin{multline}
\sum_{{\substack{\alpha\cap\beta=\emptyset\\\alpha\cup\beta\neq\emptyset}}}Q_{|\alpha|,|\beta|}\prod_{a\in\alpha}\xi_a\prod_{b\in\beta}\xi_b^2=\sum_{{p+q\geq 1}} Q_{p,q}\sum_{|\beta|=q}\prod_{b\in\beta}\xi_b^2\sum_{\mathclap{\substack{|\alpha|=p\\\alpha\cap\beta=\emptyset}}}\ \prod_{a\in\alpha}\xi_a \\
\to \sum_{{p+q\geq 1}} Q_{p,q}\sum_{|\beta|=q}\prod_{b\in\beta}\xi_b^2\sum_{\mathclap{|\alpha|=p}}\ \prod_{a\in\alpha}\xi_a \to \sum_{{p+q\geq 1}}\frac{Q_{p,q}}{q!} \left(\sum_{|\alpha|=p}\prod_{a\in\alpha}\xi_a\right)\left(\sum_b\xi_b^2\right)^q, 
\end{multline}
where each substitution is justified because of the overall constraint that allows us to neglect powers $\xi_b^{k}$ with $k\geq 3$. It can be seen that \cite{Mezard1986,abramowitz1972handbook}
\begin{equation}
 \sum_{|\alpha|=p}\prod_{a\in\alpha}\xi_a=\frac{\left(\sum_{a}\xi_a^2\right)^\frac{p}{2}}{p!}\He_p\left(\frac{\sum_a\xi_a}{\sqrt{\sum_a\xi_a^2}}\right).
\end{equation}
Here $\He_p(x)$ is the probabilists' Hermite polynomial \cite{abramowitz1972handbook}. Substituting the previous identity in the expression for $z_0$, we obtain
\begin{multline}
 z_0[\bQ]=\\
 =\int_{-\infty}^{+\infty}\frac{\dd x\dd k_x \dd y\dd k_y}{(2\pi)^2} \exp\left[\sum_{p+q\geq 1}\frac{Q_{p,q}}{p!q!}y^{\frac{p}{2}+q}\He_p\left(\frac{x}{\sqrt{y}}\right)+ik_x x+ik_y y\right]\Phi(k_x,k_y),
\end{multline}
where
\begin{equation}
 \Phi(k_x,k_y)=\left[\int_{0}^{2\pi}\frac{\dd\lambda}{2\pi}\exp\left(-2i\lambda-ik_x \e^{i\lambda}-ik_y \e^{2i\lambda}\right)\right]^n=\left(-\frac{k_x^2}{2}-ik_y\right)^n.
\end{equation}
Using now the identity $a^n=\left.\partial_t^n\e^{at}\right|_{t=0}$, we can write \cite{Mezard1985,Mezard1986,Lucibello2017}
\begin{multline}\label{zfractional}
 \frac{\ln z_0[\bQ]}{n}=\frac{1}{n}\ln\left[\left.\frac{\partial^n}{\partial t^n}\int_{-\infty}^{+\infty}\frac{\e^{-\frac{z^2}{2}}\dd z}{\sqrt{2\pi}} \exp\left(\sum_{p+q\geq 1}\frac{Q_{p,q}}{p!q!}t^{\frac{p}{2}+q}\He_p\left(z\right)\right)\right|_{t=0}\right]\\
 =\int_0^{+\infty}\frac{\dd t}{t}\left[\e^{-t}-\int_{-\infty}^{+\infty}\frac{\e^{-\frac{z^2}{2}}\dd z}{\sqrt{2\pi}} \exp\left(\sum_{p+q\geq 1}\frac{Q_{p,q}}{p!q!}(-t)^{\frac{p}{2}+q}\He_p\left(z\right)\right)\right]+O(n).
\end{multline}
At this point, some considerations are in order. Let us first observe that, on the matching saddle-point in Eq.~\eqref{matchingsp}, $Q_{p,q}^\text{sp}=\delta_{p,0}Q_q$, the previous expression becomes
\begin{equation}
  \lim_{n\to 0}\frac{\ln z_0[Q]}{n}=\int_0^{+\infty}\frac{\dd t}{t}\left[\e^{-t}- \exp\left(\sum_{q=1}^\infty\frac{Q_{q}}{q!}(-t)^{q}\right)\right].
\end{equation}
The expression above coincides with the one-site partition function in the \textsc{Rmp} and it has been discussed in details in Refs.~\cite{Mezard1985,Mezard1987,Parisi2002,Caracciolo2017}. In particular, Eqs.~\eqref{zmatchingsp1} can be obtained introducing the function $G$ in Eq.~\eqref{zmatchingsp2}. On the other hand, if we consider the \textsc{Rtsp} saddle-point solution, $\tilde Q^\text{sp}_{q,p}=\delta_{q,0}\tilde Q_p$, the \textsc{Rtsp} one-site partition function is recovered, as it can be easily seen comparing Eq.~\eqref{zfractional} with the results in Ref.~\cite{Mezard1986}.

\section{Evaluation of $\Delta S^\text{T}$ on the matching saddle-point}\label{app:dst}
To evaluate the $\Delta S^\text{T}$ contribution on the matching saddle-point, we follow the approach in Ref.~\cite{Mezard1987,Parisi2002}. In particular,
\begin{multline}
 \frac{1}{n}\Delta S^\text{T}[\bQ^\text{sp}]=\frac{1}{Nn}\sideset{}{'}\sum_{\mathclap{\alpha,\beta|\alpha,\hat\beta}}\frac{g_{|\alpha|+2|\beta|}g_{|\alpha|+2|\hat \beta|}}{4 g^2_{2|\beta|+2|\hat\beta|+2|\alpha|}}Q_{|\beta|+|\hat\beta|+|\alpha|}^2-\frac{\varepsilon-1}{4nN}\sum_{\alpha\neq\emptyset}\frac{g_{|\alpha|}}{g_{2|\alpha|}}Q_{0,|\alpha|}\\
 =\frac{1}{N}\sum_{\mathclap{\substack{s+p\geq 1\\s+q\geq 1}}}\frac{(-1)^{p+q+s-1}\Gamma(s+p+q)}{p!q!s!}\frac{g_{s+2p}g_{s+2q}}{4 g^2_{2p+2q+2s}}Q_{p+q+s}^2
 -\frac{\varepsilon-1}{4N}\sum_{p=1}^\infty\frac{(-1)^{p-1}}{p}\frac{g_{p}}{g_{2p}}Q_{p}.
\end{multline}
The asymptotic value of the last sum is given by
\begin{multline}
 \lim_{\upbeta\to +\infty}\frac{\varepsilon-1}{4N\upbeta}\sum_{p=1}^\infty\frac{(-1)^{p-1}}{p}\frac{g_{p}}{g_{2p}}Q_{p}
 =\lim_{\upbeta\to +\infty}\frac{\varepsilon-1}{2N\upbeta}\sum_{p=1}^\infty\frac{(-1)^{p-1}}{p p!}\int_{-\infty}^{+\infty}\e^{\upbeta p x-\hat G(x)}\dd x\\=\frac{\varepsilon-1}{2N}\lim_{\upbeta\to +\infty}\int_{-\infty}^{+\infty}\frac{\gamma_\text{E}+\upbeta x+\Gamma(0,\e^{\upbeta x})}{\upbeta}\e^{-\hat G(x)}\dd x\\
 =\frac{\varepsilon-1}{2N}\int_{0}^{+\infty}x\e^{-\hat G(x)}\dd x=\frac{\varepsilon-1}{4N}\zeta(2).
\end{multline}
In the previous expression, $\gamma_E$ is the Euler-Mascheroni constant, whereas $\Gamma(a,z)\coloneqq \int_z^{\infty}\e^{-t}t^{a-1}\dd t$ is the incomplete gamma function \cite{abramowitz1972handbook}. In the contribution from $\Delta S^\text{T}$ we also have
\begin{multline}
\frac{1}{n}\sum_{\substack{s+p\geq 1\\s+q\geq 1}}^\infty\binom{n}{s\ p\ q}\frac{g_{s+2p}g_{s+2q}}{4 g_{2p+2q+2s}}Q_{p+q+s}^2=\\
=\sum_{\mathclap{\substack{s+p\geq 1\\s+q\geq 1}}}\frac{(-1)^{p+q+s-1}\Gamma(s+p+q)}{p!q!s!}\frac{g_{s+2p}g_{s+2q}}{4 g^2_{2p+2q+2s}}Q_{p+q+s}^2\\
 =\sum_{\mathclap{\substack{s+p\geq 1\\s+q\geq 1}}}\frac{(-1)^{p+q+s-1}g_{s+2p}g_{s+2q}}{2p!q!s!g_{2p+2q+2s}}Q_{p+q+s}\int_{-\infty}^{+\infty}\e^{(p+q+s) x-G(x)}\dd y
 =\frac{1}{\upbeta}\int_{-\infty}^{+\infty}\dd x\, \e^{-G(x)}\\\times \sum_{k=1}^\infty (-1)^{k-1}\e^{k x}Q_k\left[\sum_{p=1}^{k-1}\frac{1}{2pp!(k-p)!}+\sum_{s=1}^k\sum_{p=0}^{k-s}\frac{1}{s!(s+2p)p!(k-s-p)!}\right].
 \end{multline}
In Ref.~\cite{Mezard1987} it has been proved that
\begin{equation}
 \lim_{\upbeta\to+\infty}\frac{1}{\upbeta^2}\int_{-\infty}^{+\infty}\dd x\, \e^{-G(x)} \sum_{k=1}^\infty (-1)^{k-1}\e^{k x}Q_k \sum_{p=1}^{k-1}\frac{1}{2pp!(k-p)!}=-\frac{\zeta(3)}{N}.
\end{equation}
The remaining contribution is zero. To prove this fact, let us start from
\begin{multline}
 \sum_{s=1}^k\sum_{p=0}^{k-s}\frac{1}{s!(s+2p)p!(k-s-p)!}=\sum_{s=1}^k\frac{i}{2\pi s!}\int_0^{+\infty}\dd t\oint_{\gamma_{\mathrm H}}\dd z \sum_{p=0}^\infty(-z)^{p+s-k-1}\frac{\e^{-(s+2p)t-z}}{p!}\\
 =\sum_{s=1}^k\frac{i}{2\pi s!}\int_0^{+\infty}\e^{-st}\dd t\oint_{\gamma_{\mathrm H}}\e^{-(1+\e^{-2t})z}(-z)^{s-k-1}\dd z
 =\sum_{s=1}^k\int_0^{+\infty}\frac{\left(1+\e^{-2t}\right)^{k-s-1}\e^{-st}}{\Gamma(k-s+1) s!}\dd t\\
 =\int_0^{1}\sum_{s=1}^k\frac{\left(1+\tau\right)^{k-s}\tau^{\sfrac{s}{2}-1}}{2s!(k-s)!}\dd\tau=\int_0^1\dd \tau\frac{\left(\tau+\sqrt \tau+1\right)^k-(1+\tau)^k}{2k!\tau}.
\end{multline}
In the expression above, $\gamma_{\mathrm H}$ is the Hankel contour in the complex plane. Summing over $k$ we get
\begin{multline}
 \frac{1}{2\upbeta}\int_{-\infty}^{+\infty}\e^{-G(x)}\dd x \int_0^1\dd \tau\frac{G\left(x+\ln\left(1+\sqrt \tau+\tau\right)\right)-G\left(x+\ln(1+\tau)\right)}{\tau}\\
 =\frac{\upbeta}{2}\int_{-\infty}^{+\infty}\e^{-\hat G(x)}\dd x \int_0^{+\infty}\dd u\left[\hat G\left(x\!+\!\frac{\ln\left(1\!+\!\e^{-\frac{\upbeta u}{2}}\!+\!\e^{-\upbeta u}\right)}{\upbeta}\right)\!-\!\hat G\left(x\!+\!\frac{\ln(1\!+\!\e^{-\upbeta u})}{\upbeta}\right)\right],\end{multline}
whose corresponding contribution goes to zero as $\upbeta\to\infty$. We can finally write
\begin{equation}
 \lim_{\upbeta\to+\infty}\lim_{n\to 0}\frac{\Delta S^\text{T}[\bQ^\text{sp}]}{n\upbeta}=-\frac{1}{N}\left(\frac{\varepsilon-1}{4}\zeta(2)+\zeta(3)\right).
\end{equation}

\section{On the evaluation of $\ln\det\boldsymbol{\mathsf \Omega}$ on the matching saddle-point}\label{app:determinante}
In this Appendix we will give some details about the evaluation of the logarithm of the determinant of the Hessian matrix $\boldsymbol\Omega$ on the matching saddle-point using the replica approach, showing that it is equal to the \textsc{Rmp} contribution, plus an additional contribution that we expect to be opposite from W{\"a}stlund's formula. We start from its general expression in Eq.~\eqref{Omega},
\begin{multline}
 \Omega_{\alpha\beta,\hat\alpha\hat\beta}[\boldsymbol{\mathsf Q}^\text{sp}]\coloneqq \sqrt{g_{|\alpha|+2|\beta|}g_{|\hat\alpha|+2|\hat\beta|}}\left.\frac{\partial^2 S[\boldsymbol{\mathsf Q}]}{\partial Q_{\alpha,\beta}\partial Q_{\hat \alpha,\hat \beta}}\right|_{\boldsymbol{\mathsf Q}=\boldsymbol{\mathsf Q}^\text{sp}}\\
 =\delta_{\alpha,\hat\alpha}\delta_{\beta,\hat\beta}\mathbb I(\alpha\cap\beta =\emptyset)
 -2\sqrt{g_{|\alpha|+2|\beta|}g_{|\hat\alpha|+2|\hat\beta|}}\left\langle\exp\left(i\sum_{\mathclap{a\in\alpha\cup\hat\alpha}}\lambda^a+2i\sum_{\mathclap{b\in\beta\cup\hat\beta}}\lambda^b\right)\right\rangle_{z_0}\\
 +2\sqrt{g_{|\alpha|+2|\beta|}g_{|\hat\alpha|+2|\hat\beta|}}\left\langle\exp\left(i\sum_ {\mathclap{a\in\alpha}}\lambda^a+2i\sum_{b\in\beta}\lambda^b\right)\right\rangle_{z_0}\left\langle\exp\left(i\sum_{a\in\hat\alpha}\lambda^a+2i\sum_{b\in\hat\beta}\lambda^b\right)\right\rangle_{z_0}\\
 =\delta_{\alpha,\hat\alpha}\delta_{\beta,\hat\beta}+\frac{1}{2}\frac{Q_{\alpha,\beta}}{\sqrt{g_{|\alpha|+2|\beta|}}}\frac{Q_{\hat\alpha,\hat\beta}}{\sqrt{g_{|\hat\alpha|+2|\hat\beta|}}}\\
 -\frac{\sqrt{g_{|\alpha|+2|\beta|}g_{|\hat\alpha|+2|\hat\beta|}}Q_{\alpha\triangle\hat\alpha,\beta\cup\hat\beta\cup(\alpha\cap\hat\alpha)}}{g_{|\alpha\triangle\hat\alpha|+2|\beta\cup\hat\beta\cup(\alpha\cap\hat\alpha)|}}\mathbb I(\beta\cap\hat\beta=\emptyset)\mathbb I((\alpha\cup\hat\alpha)\cap(\beta\cup\hat\beta)=\emptyset).\end{multline}
In the replica symmetric ansatz on the matching saddle-point solution the expression greatly simplifies, becoming
\begin{multline}
 \Omega_{\alpha\beta,\hat\alpha\hat\beta}=\delta_{\alpha,\hat\alpha}\delta_{\beta,\hat\beta}\mathbb I(\beta\cap \alpha=\emptyset)+\frac{\delta_{|\alpha|,0}\delta_{|\hat \alpha|,0}}{2}\frac{Q_{|\beta|}}{\sqrt{g_{2|\beta|}}}\frac{Q_{|\hat\beta|}}{\sqrt{g_{2|\hat\beta|}}}\\
 -\delta_{\alpha,\hat\alpha}\frac{Q_{|\beta\cup\hat\beta\cup\alpha|}\sqrt{g_{|\alpha|+2|\beta|}g_{|\alpha|+2|\hat\beta|}}\mathbb I(\beta\cap\hat\beta=\emptyset)\mathbb I((\beta\cup\hat\beta)\cap \alpha=\emptyset)}{g_{2|\beta\cup\hat\beta\cup\alpha|}}.\end{multline}
We follow the approach of Refs.~\cite{Mezard1987,Parisi2002}. Observe first that the quantity above is diagonal respect to the index $\alpha$. In particular
\begin{multline}
 \ln\det\boldsymbol{\mathsf \Omega}[\boldsymbol{\mathsf Q}^\text{sp}]=\ln\det\boldsymbol{\mathsf \Omega}^{(0)}[\boldsymbol{\mathsf Q}^\text{sp}]+\sum_{|\alpha|=1}^{\infty}\binom{n}{|\alpha|}\ln\det\boldsymbol{\mathsf \Omega}^{(|\alpha|)}[\boldsymbol{\mathsf Q}^\text{sp}]\\
 =\ln\det\boldsymbol{\mathsf \Omega}^{(0)}[\boldsymbol{\mathsf Q}^\text{sp}]+n\sum_{s=1}^{\infty}\frac{(-1)^{s-1}}{s}\ln\det\boldsymbol{\mathsf \Omega}^{(s)}[\boldsymbol{\mathsf Q}^\text{sp}]+o(n),
\end{multline}
where we have separated the contributions for different values of $s=|\alpha|$ and introduced
\begin{equation}
\Omega^{(s)}_{\beta\hat\beta}\coloneqq \delta_{\beta,\hat\beta}-Q_{|\beta|+|\hat\beta|+s}\mathbb I(\beta\cap\hat\beta=\emptyset)\frac{\sqrt{g_{2|\beta|+s}g_{2|\hat\beta|+s}}}{g_{2|\beta|+2|\hat\beta|+2s}}+\frac{\delta_{s,0}}{2}\frac{Q_{|\beta|}Q_{|\hat\beta|}}{\sqrt{g_{2|\beta|}g_{2|\hat\beta|}}},
\end{equation}
where $\beta$ is a set of $n-s$ replica indices. A vector $\boldsymbol{\mathsf q}$ is eigenvector of $\boldsymbol{\mathsf \Omega}^{(s)}$ with eigenvalue $\lambda$ if
\begin{equation}\label{eigeneq}
 q_\beta-\sum_{\mathclap{\hat\beta\colon \beta\cap\hat\beta=\emptyset}}\frac{\sqrt{g_{s+2|\beta|}g_{s+2|\hat\beta|}}}{g_{2|\beta|+2|\hat\beta|+2s}}Q_{|\beta|+|\hat\beta|+s}q_{\hat \beta}
 +\frac{\delta_{s,0}}{2}\sum_{|\hat\beta|\neq\emptyset}\frac{Q_{|\beta|}Q_{|\hat\beta|}q_{\hat\beta}}{\sqrt{g_{2|\beta|}g_{2|\hat\beta|}}}=\lambda q_\beta.
\end{equation}
A first step is to diagonalize $\boldsymbol{\mathsf \Omega}^{(s)}$ according to the irreducible representations of the permutation group \cite{Wignergroup} in the space of $n-s$ replica indices. In the spirit of the strategy of De Almeida and Thouless \cite{DeAlmeida1978}, and strictly following Refs.~\cite{Mezard1987,Parisi2002}, we observe that an eigenvector $\boldsymbol{\mathsf q}^{(c)}$ with $c$ distinguished replicas, is such that
\begin{equation}
 q_\beta^{(c)}\coloneqq\begin{cases}
                    0&\text{if }|\beta|<c,\\
                    \omega_{|\beta|}^i&\text{if $\beta$ contains $c-i$ of the $c$ distinguished indices, with $i=0,1,\dots,c$.}
                   \end{cases}
\end{equation}
For $c=0$, than we only have $n-s$ possible eigenvectors in the form $q_\beta^{(0)}\equiv q^{(0)}_{|\beta|}$. The eigenvalue equation can be written down for the $(n-s)\times (n-s)$ matrix $\boldsymbol{\mathsf N}^{(s,0)}$ given by
\begin{equation}
 N^{(s,0)}_{pq}=\delta_{pq}-\binom{n-s-p}{q}\frac{\sqrt{g_{2p+s}g_{2q+s}}Q_{p+q+s}}{g_{2p+2q+2s}}
 +\frac{\delta_{s,0}}{2}\binom{n}{q}\frac{Q_pQ_q}{\sqrt{g_{2p}g_{2q}}},
\end{equation}
whose eigenvalues have multiplicity $1$ in the set of eigenvalues of $\boldsymbol{\mathsf \Omega}^{(s)}$. For $c\geq 1$, imposing the ortogonality relation between $\boldsymbol{\mathsf q}^{(c)}$ and $\boldsymbol{\mathsf q}^{(c-1)}$, we obtain 
\begin{subequations}
\begin{multline}
 \sum_{\beta}q^{(c)}_\beta q^{(c-1)}_\beta=\sum_{|\beta|\geq c}q^{(c)}_\beta q^{(c-1)}_\beta\\
 =\sum_{p\geq c}\sum_{j=0}^{c-1}\sum_{r=0}^{c-j}\binom{n-s-c}{p-j-r}\binom{c}{j}\binom{c-j}{r}\omega_{p}^{c-(r+j)}\omega_{p}^{c-1-j}=0\\
 \Longrightarrow \sum_{r=0}^{c-j}\binom{n-s-c}{p-j-r}\binom{c-j}{r}\omega_{p}^{c-(r+j)}=0\text{ for $p\geq c$ and $j=0,1,\dots,c-1$},\label{orto}
\end{multline}
that for $n\to 0$ becomes, for $p\geq c$ and $j=0,1,\dots,c-1$,
\begin{equation}\sum_{r=0}^{c-j}(-1)^{r}\binom{c-j}{r}\frac{\Gamma(s+c+p-r-j)}{\Gamma(p-r-j+1)}\omega_{p}^{c-(r+j)}=0,\label{orto2}
\end{equation}
\end{subequations}
where we have used the property
\begin{equation}
 \lim_{n\to 0}\binom{n-a}{b}=\frac{(-1)^b\Gamma(a+b)}{\Gamma(a)\Gamma(b+1)}.
\end{equation}
Eq.~\eqref{orto} allows us to keep $\omega_{p}^0$ as independent only. In particular, for $c=1$ we have $p\omega_{p}^0+(n-s-p)\omega_{p}^1=0$ and therefore the diagonalization of $\boldsymbol{\mathsf \Omega}$ in the subspace $c=1$ can be reduced to the diagonalization of the $(n-s-1)\times(n-s-1)$ matrix
\begin{multline}
 N^{(s,1)}_{pq}=\delta_{pq}
 -\binom{n-s-p}{q}\frac{q}{q+s-n}\frac{\sqrt{g_{2p+s}g_{2q+s}}Q_{p+q+s}}{g_{2p+2q+2s}}\\
 +\frac{\delta_{s,0}}{2}\left[\binom{n-1}{q}\frac{q}{q-n}+\binom{n-1}{q-1}\right]\frac{Q_pQ_q}{\sqrt{g_{2p}g_{2q}}},
\end{multline}
with eigenvalue multiplicity $n-s-1$ respect to the original matrix $\boldsymbol{\mathsf \Omega}^{(s)}$. Before proceeding further, some considerations are in order. As observed in Refs.~\cite{Mezard1987,Parisi2002}, the matrices $\boldsymbol{\mathsf N}^{(0,0)}$ and $\boldsymbol{\mathsf N}^{(0,1)}$ have the same limit as $n\to 0$, in particular $\lim_{n\to 0}\boldsymbol{\mathsf N}^{(0,1)}=\boldsymbol{\mathsf N}^{(0,0)}$. The calculation of the contribution of these two matrices requires some care, but it can be proved that it is eventually zero for $\beta\to+\infty$ \cite{Mezard1987,Parisi2002}. For $c\geq 1$ Eq.~\eqref{orto2} implies
\begin{multline}
 \frac{\omega_{p}^0}{\prod_{u=0}^{c-1}(s+p+u)}=\frac{\omega_{p}^1}{(p-c+1)\prod_{u=1}^{c-1}(s+p+u)} =\cdots\\
 =\cdots=\frac{\omega_{p}^l}{\prod_{v=1}^{l}(p-c+v)\prod_{u=l}^{c-1}(s+p+u)}
 =\cdots =\frac{\omega_{p}^{c}}{\prod_{v=1}^{c}(p-c+v)}\\
 \Longrightarrow \omega^l_p=\frac{\Gamma(p+l-c+1)\Gamma(s+p)}{\Gamma(p-c+1)\Gamma(s+p+l)}\omega_p^0.
\end{multline}
Using the previous result, Eq.~\eqref{eigeneq} for the $|\beta|=0$ component of an eigenvector with $c\geq1$ becomes
\begin{multline}
 \lambda\omega_p^0=\frac{\delta_{s,0}}{2}\sum_{q=1}^\infty\sum_{i=0}^c\binom{n-c}{q+i-c}\binom{c}{c-i}\frac{Q_pQ_q\omega_q^i}{\sqrt{g_{2p}g_{2q}}}\\
+\omega_p^0-\sum_{q=1}^\infty\binom{n-s-p}{q}\frac{\sqrt{g_{s+2p}g_{s+2q}}}{g_{2s+2p+2q}}Q_{p+q+s}\omega_{q}^c\\
 =\omega_p^0-\sum_{q=1}^\infty (-1)^{q}\frac{\Gamma(s+q)\Gamma(s+p+q)\sqrt{g_{s+2p}g_{s+2q}}}{\Gamma(s+p)\Gamma(s+q+c)\Gamma(q-c+1)}\frac{Q_{p+q+s}}{g_{2s+2p+2q}}\omega_{q}^0+o(n)
\end{multline}
(observe that the quadratic term is zero because of Eq.~\eqref{orto}). We can finally write for each value of $s$
\begin{equation}
 \ln\det\boldsymbol{\mathsf \Omega}^{(s)}=\sum_{c=0}^\infty\left[\binom{n-s}{c}-\binom{n-s}{c-1}\right]\ln\det\boldsymbol{\mathsf N}^{(s,c)}
\end{equation}
being the $(n-s-c)\times(n-s-c)$ matrix $\boldsymbol{\mathsf N}^{(s,c)}$ in the $n\to 0$ limit
\begin{equation}
 N^{(s,c)}_{pq}=\delta_{pq}
 -(-1)^q\frac{\Gamma(s+q)\Gamma(p+q+s)\sqrt{g_{2p+s}g_{2q+s}}}{\Gamma(s+p)\Gamma(q-c+1)\Gamma(s+q+c)}\frac{Q_{p+q+s}}{g_{2p+2q+2s}}.
\end{equation}
Shifting by $c$ the indices in the expression above, transposing and then multiplying by \[(-1)^{p+q}\sqrt{\frac{g_{2q+2c+s}}{g_{2p+2c+s}}}\frac{\Gamma(q+c+s)\Gamma(p+1)}{\Gamma(p+c+s)\Gamma(q+1)}\] we get a new matrix $\boldsymbol{\mathsf M}^{(s,c)}$ with the same spectrum of $\boldsymbol{\mathsf N}^{(s,c)}$, namely
\begin{equation}
 M^{(s,c)}_{pq}=\delta_{pq}
 -(-1)^{q+c}\frac{\Gamma(p+q+s+2c)}{\Gamma(p+s+2c)\Gamma(q+1)}\frac{g_{2q+2c+s}}{g_{2(p+q+s+2c)}}Q_{p+q+s+2c}.
\end{equation}
To evaluate the correction to the free-energy we need to evaluate therefore
\begin{multline}\label{omegamsc}
\ln\det\boldsymbol{\mathsf \Omega}=\\
\begin{split}
=&\sum_{c=2}^\infty\left[\binom{n}{c}-\binom{n}{c-1}\right]\ln\det\boldsymbol{\mathsf M}^{(0,c)}+\sum_{s=1}^\infty\binom{n}{s}\sum_{c=0}^\infty\left[\binom{n-s}{c}\!-\!\binom{n-s}{c-1}\right]\ln\det\boldsymbol{\mathsf M}^{(s,c)}\\
=&n\sum_{c=2}^\infty(-1)^{c-1}\frac{2c-1}{c(c-1)}\ln\det\boldsymbol{\mathsf M}^{(0,c)}\\
&-n\sum_{s=1}^\infty
\sum_{c=0}^\infty (-1)^{c+s}(2c+s-1)\frac{\Gamma(s+c-1)}{s!c!}\ln\det\boldsymbol{\mathsf M}^{(s,c)}+o(n).
\end{split}
\end{multline}
At this point is important to observe that, in the expression above, the $s=0$ contribution exactly coincides with the fluctuation contribution appearing in the finite-size corrections of \textsc{Rmp}. Its evaluation has been performed in Refs.~\cite{Mezard1987,Parisi2002}, but no closed formula is known for it. The $s\geq1$ contribution is instead absent in the \textsc{Rmp}. Generalizing therefore the analysis of Refs.~\cite{Mezard1987,Parisi2002}, we note that eigenvalues of $\boldsymbol{\mathsf M}^{(s,c)}$ are the same as the eigenvalues of the operator
\begin{equation}
\mathcal M^{(s,c)}(x,y)=\delta(x-y)-(-1)^c\mathcal A^{(c+\sfrac{s}{2})}(x,y)
\end{equation}
where the operator $\mathcal A^{(k)}(x,y)$ was the one introduced in Ref.~\cite{Parisi2002} and it is defined as
\begin{equation}
\mathcal A^{(k)}(x,y)=2\e^{-\frac{G(x)+G(y)}{2}}\sum_{q=0}^\infty \frac{(-1)^q\e^{(q+k)(x+y)}}{\Gamma(2k+q)\Gamma(q+1)}g_{2(q+k)}.
\end{equation}
Indeed, if $\psi_p$ is an eigenvector of $\boldsymbol{\mathsf M}^{(s,c)}$ with corresponding eigenvalue $\lambda$, then, by a straightforward computation, it can be verified that
\begin{equation}
 \phi(x)\coloneqq\e^{\left(c+\frac{s}{2}\right)x-\frac{G(x)}{2}}\sum_{p=0}^{+\infty}\frac{(-1)^p}{p!}{g_{2p+2c+s}}\psi_p\e^{px}
\end{equation}
is an eigenvector of $\mathcal M^{(s,c)}(x,y)$ with the same eigenvalue.
To evaluate the $\upbeta\to\infty$ limit of $\mathcal A^{(k)}(x,y)$ we observe that it has the same eigenvalues of the operator
\begin{equation}
\mathcal H^{(k)}(x,y)\coloneqq \upbeta \mathcal A^{(k)}(\upbeta x,\upbeta y)
 =\e^{-\frac{\hat G(x)+\hat G(y)}{2}}\sum_{q=0}^\infty \frac{(-1)^q\e^{(q+k)\upbeta (x+y)}}{\Gamma(2k+q)\Gamma(q+1)}\frac{1}{q+k}.
\end{equation}
In Ref.~\cite{Mezard1987,Parisi2002} it has been shown that, for $\upbeta\to+\infty$, if we impose $\ln k=\upbeta t$ with $t$ fixed, the $\upbeta\to+\infty$ limit exists. In particular
\begin{equation}
 \mathcal H^{(k)}(x,y)\xrightarrow[\text{$t$ fixed}]{\ln k=\upbeta t} \mathcal H_{t}(x,y)=\e^{-\frac{\hat G(x)+\hat G(y)}{2}}\theta\left(x+y-2t\right).
\end{equation}
This result suggests that the evaluation of the sums in Eq.~\eqref{omegamsc} must be performed scaling $c$, $s$ and $\upbeta$ in a proper way. Given the known result for the \textsc{Rmp}, we distinguish now between the contributions with $s\geq 1$ and the contribution obtained for $s=0$. We know indeed that we can obtain a finite limit for the $s=0$, that is the corresponding fluctuation correction to the \textsc{aoc} in the \textsc{Rmp}. We can write
\begin{multline}\label{detrmp}
n\sum_{c=2}^\infty(-1)^{c-1}\frac{2c-1}{c(c-1)}\ln\det\mathbf M^{(0,c)}=\\
\begin{split}
&=n\sum_{E=1}^\infty \frac{1}{E}\sum_{c=2}^\infty(-1)^{(E+1)c}\frac{2c-1}{c(c-1)}\mathrm{tr}\left[\left(\mathcal A^{(c)}\right)^E\right]\\
&=n\sum_{E=1}^\infty\frac{1}{E} \sum_{c=1}^\infty\frac{(4c-1)\mathrm{tr}\left[\left(\mathcal A^{(2c)}\right)^E\right]}{2c(2c-1)}
-n\sum_{E=1}^\infty\frac{(-1)^E}{E}\sum_{c=1}^\infty\frac{(4c+1)\mathrm{tr}\left[\left(\mathcal A^{(2c+1)}\right)^E\right]}{2c(2c+1)}\\
&\xrightarrow{\upbeta\to +\infty}2n\upbeta\sum_{E\text{ odd}}\frac{1}{E}\int_0^{+\infty}\mathrm{tr}\left[\mathcal H^E_t\right]\dd t.
\end{split}
\end{multline}
A numerical estimation of the quantity above can be found in Refs.~\cite{Parisi2002,Lucibello2017}. The evaluation of the $s\geq 1$ terms is more complicated and we will present here a non-rigorous treatment. We have
\begin{multline}\label{settoriE}
n\sum_{s=1}^\infty\sum_{c=0}^\infty (-1)^{c+s-1}(2c+s-1)\frac{\Gamma(s+c-1)}{s!c!}\ln\det\boldsymbol{\mathsf M}^{(s,c)}\\
\begin{split}
=&n\sum_{E=1}^\infty\frac{1}{E}\sum_{s=1}^\infty\sum_{c=0}^\infty (-1)^{(E+1)c+s}(2c+s-1)\frac{\Gamma(s+c-1)}{s!c!}\mathrm{tr}\left[\left(\mathcal A^{(c+\frac{s}{2})}\right)^E\right]+o(n)\\
=& n\sum_{E=1}^\infty\frac{1}{E}\sum_{s=1}^\infty\sum_{c=0}^\infty (-1)^{s}(4c+s-1)\frac{\Gamma(s+2c-1)\mathrm{tr}\left[\left(\mathcal A^{(2c+\frac{s}{2})}\right)^E\right]}{s!(2c)!}\\ &-n(-1)^{E}\sum_{E=1}^\infty\frac{1}{E}\sum_{s=1}^\infty\sum_{c=0}^\infty (-1)^{s}(4c+s+1)\frac{\Gamma(s+2c)\mathrm{tr}\left[\left(\mathcal A^{(2c+1+\frac{s}{2})}\right)^E\right]}{s!(2c+1)!}+o(n).
\end{split}
\end{multline}
Introducing $4c+s=z$, the first sum in the last line of Eq.~\eqref{settoriE} becomes
\begin{subequations}
\begin{multline}
\sum_{s=1}^\infty\sum_{c=0}^\infty (-1)^{s}(4c+s-1)\frac{\Gamma(s+2c-1)\mathrm{tr}\left[\left(\mathcal A^{(2c+\frac{s}{2})}\right)^E\right]}{s!(2c)!}\\
=\sum_{z=1}^\infty (z-1)\mathrm{tr}\left[\left(\mathcal A^{(\sfrac{z}{2})}\right)^E\right]\sum_{s=1}^\infty\sum_{c=0}^\infty \frac{(-1)^{s}\Gamma\left(s+2c-1\right)\mathbb I\left(s+4c=z\right)}{\Gamma(s+1)\Gamma\left(2c+1\right)}
=\mathrm{tr}\left[\left(\mathcal A^{(1)}\right)^E\right]\\
+\sum_{z=3}^\infty (z-1)\mathrm{tr}\left[\left(\mathcal A^{(\sfrac{z}{2})}\right)^E\right]\sum_{s=1}^\infty\sum_{c=0}^\infty \frac{(-1)^{s}\Gamma\left(s+2c-1\right)\mathbb I\left(s+4c=z\right)}{\Gamma(s+1)\Gamma\left(2c+1\right)}.
\end{multline}
Similarly, after the change of variable $z=4c+s+2$, the second sum becomes
\begin{multline}
\sum_{s=1}^\infty\sum_{c=0}^\infty (-1)^{s}(4c+s+1)\frac{\Gamma(s+2c)}{s!\Gamma(2c+2)}\tr\left[\left(\mathcal A^{(2c+\frac{s}{2}+1)}\right)^E\right]\\
=\sum_{z=3}^\infty (z-1)\tr\left[\left(\mathcal A^{(\sfrac{z}{2})}\right)^E\right]\sum_{s=1}^\infty\sum_{c=0}^\infty \frac{(-1)^{s}\Gamma(s+2c)\mathbb I\left(2+s+4c=z\right)}{\Gamma(s+1)\Gamma(2c+2)}.
\end{multline}
\end{subequations}
In other words, we can write
\begin{multline}
 \sum_{s=1}^\infty\sum_{c=0}^\infty (-1)^{c+s-1}(2c+s-1)\frac{\Gamma(s+c-1)}{s!c!}\ln\det\boldsymbol{\mathsf M}^{(s,c)}\\=
 \sum_{E=1}^\infty\frac{1}{E}\sum_{z=3}^\infty (z-1)\tr\left[\left(\mathcal A^{(\sfrac{z}{2})}\right)^E\right]\left[h_1(z)-(-1)^E h_2(z)\right]
\end{multline}
The main difficulty in the evaluation of the quantities above is that the coefficients
\begin{align}
h_1(z)&\coloneqq\sum_{s=1}^\infty\sum_{c=0}^\infty \frac{(-1)^{s}\Gamma\left(s+2c-1\right)\mathbb I\left(s+4c=z\right)}{\Gamma(s+1)\Gamma\left(2c+1\right)}\\
h_2(z)&\coloneqq\sum_{s=1}^\infty\sum_{c=0}^\infty \frac{(-1)^{s}\Gamma(s+2c)\mathbb I\left(2+s+4c=z\right)}{\Gamma(s+1)\Gamma(2c+2)}
\end{align}
are oscillating with diverging amplitude in $z$ for $z\to +\infty$, and therefore the large $z$ estimation is not straightforward. It is possible, however, that a different rearrangement of the contributions appearing in the sums might lead to a simpler asymptotic evaluation.

In the present work, we avoided this estimation using W{\"a}stlund's formula, but it is interesting to observe that Eq.~\eqref{wastformula} implies that the contribution from Eq.~\eqref{settoriE} is equal and opposite to the one in Eq.~\eqref{detrmp}, that coincides with the fluctuation finite-size correction to the \textsc{aoc} in the \textsc{Rmp}. A more comprehensive study of these quantities can be matter of future investigations.

\section*{References}
\bibliographystyle{iopart-num}
\bibliography{biblio.bib}
\end{document}